\documentclass[amsmath,twocolumn,showpacs,amssymb,aps,nofootinbib]{revtex4-2}
\usepackage{graphicx}
\usepackage{subcaption}
\usepackage{pgfplots}
\usepackage{dcolumn}
\usepackage{bm}
\usepackage{slashed}
\usepackage{amsmath}
\usepackage{amssymb}
\usepackage{tikz}
\usepackage{bbold}
\usepackage{mathtools}
\usepackage{romannum}
\usepackage{natbib}
\usepackage{braket}
\usepackage{appendix}
\begin{document}
\pagenumbering{arabic}
\title{Complexity and quenches in models with three and four spin interactions \\}
\author{Mamta Gautam}
\email{mamtag@iitk.ac.in}
\author{Nitesh Jaiswal}
\email{nitesh@iitk.ac.in}
\author{Ankit Gill}
\email{ankitgill20@iitk.ac.in}
\author{Tapobrata Sarkar}
\email{tapo@iitk.ac.in}
\affiliation{
Department of Physics, Indian Institute of Technology Kanpur-208016, India}
\date{\today}
\begin{abstract}
We study information theoretic quantities in models with three and four spin interactions. These 
models show distinctive characteristics compared to their nearest 
neighbour counterparts. Here, we quantify these in terms of the Nielsen complexity in static and quench 
scenarios, the Fubini-Study complexity, and the entanglement entropy. The models that we study have a rich phase
structure, and we show how the difference in the nature of phase transitions in these, compared to ones
with nearest neighbour interactions, result in different behaviour of information theoretic quantities,
from ones known in the literature. For example, the derivative of the Nielsen complexity does not diverge 
but shows a discontinuity near continuous phase transitions, and the Fubini-Study complexity may be
regular and continuous across such transitions.
The entanglement entropy shows a novel discontinuity both at first and second order quantum phase transitions. 
We also study multiple quench scenarios in these models and contrast these with quenches in the transverse XY model. 

\end{abstract}
\maketitle
\section{\label{sec:level1}Introduction}

Over the last few years, study of information theory in quantum many body systems have fast gained 
popularity. A primary reason for this is that apart from being the foundations of what is perhaps the technology
of the future, such studies point to the deep connection between diverse areas of physics. Indeed, various
information theoretic measures such as the quantum information metric (QIM) \cite{zanardi, gu, polkov} 
and its associated Fubini-Study complexity (FSC) \cite{Chapman}, Nielsen complexity (NC) \cite{Nielsen, Nielsen1}, 
Loschmidt echo (LE) \cite{Peres, Hamma1} and entanglement entropy (EE) \cite{osborne, oster, vidal, su}, 
find extensive applications in different areas of physics, from statistical systems to black holes, the connections 
often being realised in quantum field theory via the gauge-gravity duality 
\cite{Susskind1, Susskind2, Susskind3, Susskind4, Myers1, BSS, Myers2, KKS, HM, ABHKM}.

Spin systems provide an ideal laboratory for studying information theory in a quantum mechanical setting, primarily because
these are often amenable to analytical results and provide deep physical insights regarding the behaviour of
information geometry near zero temperature quantum phase transitions (QPTs). As of now, these are well studied in,
for example, the transverse XY model, a quasi-free fermionic model with nearest neighbour (NN) interactions \cite{tapo,tapo1}, 
and the Lipkin-Meshkov-Glick model, which is an exactly solvable model with infinite range interactions \cite{tapo2,GGCHV}, 
in both time-independent and time-dependent (quench) scenarios. The purpose of the present paper is to go beyond the physics of nearest
neighbour interactions and study these information theoretic quantities in spin models in the presence of 
three and four spin interactions. Fortunately, an exactly solvable model with such interactions is known, and was 
discussed in \cite{Zvyagin} (see also \cite{Zvyagin1, Zvyaginbook}). Here, we study this model with some specific 
simplifying choice of parameters (that nonetheless preserves a rich phase structure), and show that there are some
remarkable differences (as well as similarities) in the behaviour of information theoretic quantities when 
short range interactions are added to the NN one. 

The motivation for this study is twofold. Firstly, to the best of our knowledge, short range interactions 
apart from the ubiquitous NN one and the infinite range Lipkin-Meshkov-Glick model have not 
been studied in the current literature on spin models. As we show in sequel, our model shows several distinctive
features as far as information theoretic quantities are concerned. Secondly, the nature of the phase 
transitions in the model of \cite{Zvyagin} are very different from the ones in say, the transverse XY model,
and it is of interest to understand the behaviour of information theory in this context, and we show in sequel
that some of the well known features of models with NN interactions are modified here. 

With these motivations, in this paper, we first compute the QIM and the structure of geodesics for all the 
phases of the theory in the presence of four spin interactions. Our analysis indicates that contrary to some known examples, 
the Ricci scalar of the QIM and the FSC do not capture the entire phase structure of the theory. Interestingly,
these show their typical behaviours only at a particular type of second order phase transition, namely the gapped ferromagnetic
to the gapless ferrimagnetic phases, and not between two ferrimagnetic phases (with different
magnetisations). Next, we compute the NC for this model and show that its derivative does not diverge at second order 
phase transitions, but rather shows a discontinuity, again in contrast to known examples in the literature. 
A quench scenario is considered thereafter, where we couple our model to a spin half system, and compute the
LE. Thereafter, using the density matrix renormalisation group methods \cite{white, Scholl, osborne1, gu}, we
compute the entanglement entropy, and find sharp jumps across both first order and second order phase transitions,
a result that is in contradiction to the ones known in the literature, namely that the EE shows discontinuity at a 
first-order QPT, while a cusp or a kink in the EE indicates the second-order QPT \cite{deger, huang}. Finally,
we also consider multiple global quench scenarios in the model and contrast this with such quenches in the 
transverse XY model. 

In the following sections \ref{model} to \ref{entanglement}, we elaborate upon the above points, and the paper ends with our
conclusions in section \ref{conclusions}. This paper also includes two appendices \ref{AppA} and \ref{AppB}
where some mathematical details are relegated to. Units with $\hbar = 1$ are used throughout. 

\section{\label{model}The Model}
A quantum spin models with alternating NN couplings, and three spin and four spin 
exchange interactions was studied by \cite{Zvyagin}. This will be used extensively in the paper, 
and we first recall their main results. More details can be found in \cite{Zvyagin1, Zvyaginbook}. 
The model Hamiltonian is given by
\begin{eqnarray}
{\cal H } &=&-H\!\!\sum\limits_{n}\!\!\left(\mu_{1}S_{n,1}^{z}\!+\!\mu_{2}S_{n,2}^{z}\right)\!-\!J_{1}
\!\!\sum\limits_{n}\!\!\left(S_{n,1}^{x}S_{n,2}^{x}\!+\!S_{n,1}^{y}\!S_{n,2}^{y}\right)\notag\\
& &
-J_{2}\sum\limits_{n}\left(S_{n,2}^{x}S_{n+1,1}^{x}+S_{n,2}^{y}S_{n+1,1}^{y}\right)\notag\\ 
& &-J_{13}\sum\limits_{n}\left(S_{n,1}^{x}S_{n,2}^{z}S_{n+1,1}^{x}+S_{n,1}^{y}S_{n,2}^{z}S_{n+1,1}^{y}\right)\notag\\
& &-J_{23}\sum\limits_{n}\left(S_{n,2}^{x}S_{n+1,1}^{z}S_{n+1,2}^{x}+S_{n,2}^{y}S_{n+1,1}^{z}S_{n+1,2}^{y}\right)
\notag\\& &-\!J_{14}\!\!\sum\limits_{n}\!\!\left(\!S_{n,1}^{x}\!S_{n,2}^{z}\!S_{n+1,1}^{z}\!S_{n+1,2}^{x}\!+\!S_{n,1}^{y}
\!S_{n,2}^{z}\!S_{n+1,1}^{z}\!S_{n+1,2}^{y}\!\right)\notag\\& &-J_{24}\sum\limits_{n}
\big(S_{n,2}^{x}S_{n+1,1}^{z}S_{n+1,2}^{z}S_{n+2,1}^{x}+\notag\\& &\qquad\qquad\;\; S_{n,2}^{y}
S_{n+1,1}^{z}S_{n+1,2}^{z}S_{n+2,1}^{y}\big)~.\label{Hamilt}
\end{eqnarray} 
Here, the indices $1$ and $2$ indicate two sub-lattices with nearest neighbour spins occupying two different
sub-lattices (the total number of cells will be denoted by $N$ in what follows,
where we will assume $N$ to be odd) and $S_{n,1,2}^{x,y,z}$ are the spin-$\frac{1}{2}$ 
operators for sub-lattices $1$ and $2$ at the $n$-th site. Also, $\mu_{1,2}$ are the Bohr magnetons in sub-lattices 
$1$ and $2$, $H$ is the external magnetic field along the $z$ axis, $J_{1,2}$ 
are the alternating coupling constants between nearest-neighbour spins, $J_{13,23}$ are the alternating 
coupling constants for three-spin interaction, and  similarly, $J_{14,24}$ are the alternating coupling constants for 
four-spin interactions. 

The large number of interaction parameters make a general analysis of the model cumbersome even if one chooses an
overall scale, and we will make some simplifying choices. For ease of computation, we will in this paper, set $J_{24}=0$, 
and it is further convenient to re-parametrise the Hamiltonian in terms of new couplings, $h$, $J$, and $J_{3}$ to get
rid of some unwanted fractions, as : 
\begin{center}
\(\mu_{1}=3\mu\),\space\(\mu_{2}=\mu\),\space\(\mu H=h/2\),\space\space \(J_{1}=2J\),
\space \(J_{2}=-1\),\\ \(J_{13}=5J_{3}\),\space\(J_{23}=J_{3}\),\space\space and \(J_{14}=4\)~. 
\end{center}
Then the Hamiltonian Eq. (\ref{Hamilt}) in terms of these new couplings is 
\begin{eqnarray}              
{\cal H}&=&-\frac{h}{2}\sum\limits_{n}\left(3S_{n,1} ^{z}+S_{n,2}^{z}\right)-2J\sum\limits_{n}
\big(S_{n,1}^{x}S_{n,2}^{x}+\notag\\& &S_{n,1}^{y}S_{n,2}^{y}\big)+\sum\limits_{n}
\big(S_{n,2}^{x}S_{n+1,1}^{x}+S_{n,2}^{y}S_{n+1,1}^{y}\big)\notag\\& &
-J_{3}\sum\limits_{n}\big(5S_{n,1}^{x}S_{n,2}^{z}S_{n+1,1}^{x}+5S_{n,1}^{y}S_{n,2}^{z}S_{n+1,1}^{y}\notag\\& 
&+S_{n,2}^{x}S_{n+1,1}^{z}S_{n+1,2} ^{x}+S_{n,2}^{y}S_{n+1,1}^{z}S_{n+1,2}^{y}\big)-\notag\\& &
4\!\!\sum\limits_{n}\!\big(S_{n,1}^{x}S_{n,2}^{z}S_{n+1,1}^{z}\!S_{n+1,2}^{x}+S_{n,1}^{y}S_{n,2}^{z}
S_{n+1,1}^{z}\!S_{n+1,2} ^{y}\big)~.\label{parHamilt}\notag\\
\end{eqnarray}        
The Hamiltonian Eq. (\ref{parHamilt}) can be diagonalised, and dispersion relations can be 
obtained by using the Jordan-Wigner, Fourier and Bogoliubov transformations detailed in Appendix \ref{AppA}, 
and after these transformations, we finally obtain
\begin{equation}
{\cal H}=\sum\limits_{k}\left({\cal E}_{k,1}b_{k,1}^{\dagger}b_{k,1}+{\cal E}_{k,2}b_{k,2}^{\dagger}b_{k,2}\right)-Nh~,
\end{equation}
with the dispersion relation,
\begin{equation}
{\cal E}_{k,1,2}=\left(h-\frac{3J_{3}}{2}\cos k\right)\mp\Lambda_{k}~.\label{energy}
\end{equation}   
%%%%%%%%%%%%%%%%%%%%%%%%%%%%%%%%%%%%%%%%%%%%%%%
\begin{figure}[h!]
\centering
\includegraphics[width=0.47\textwidth]{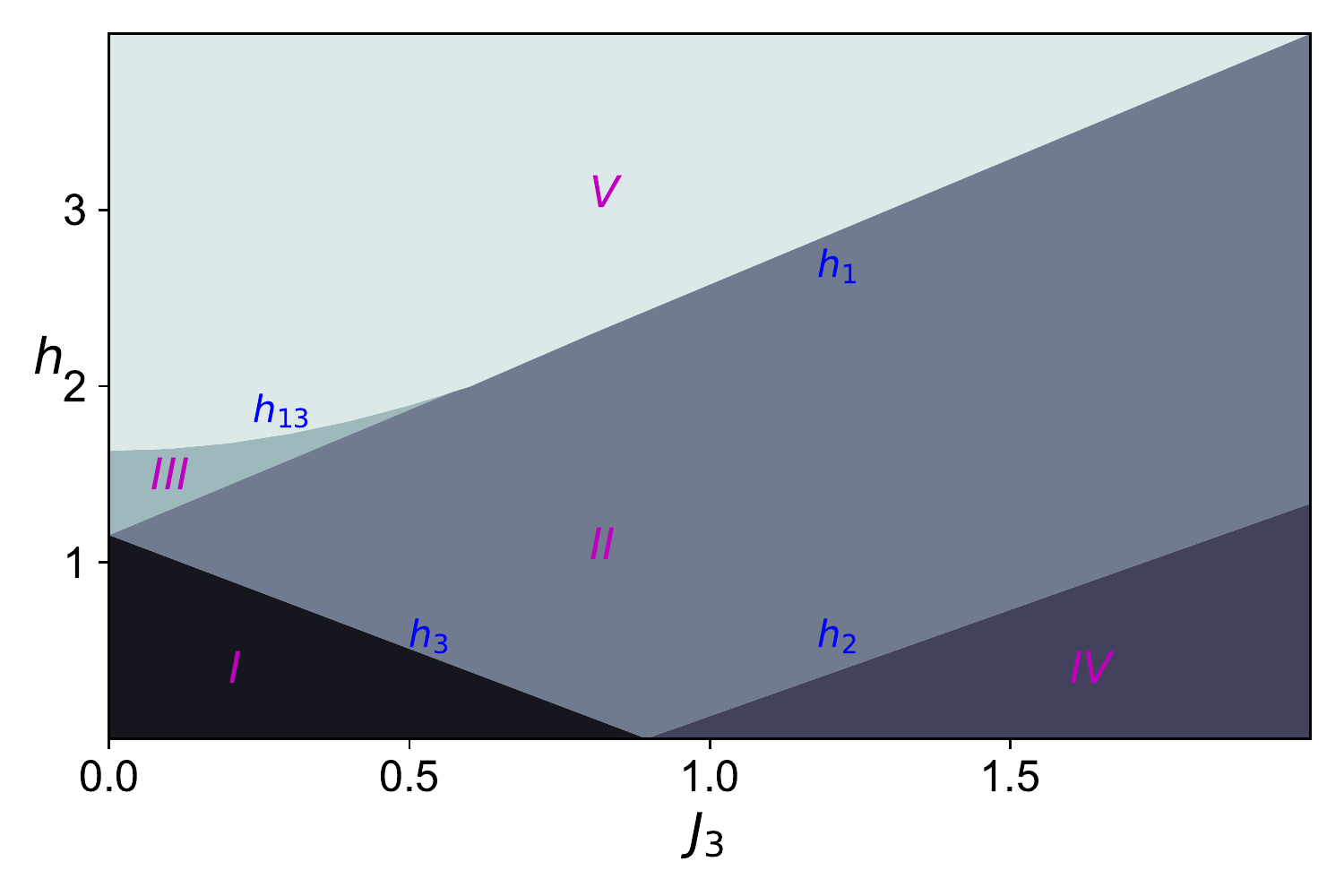}
\caption{The phase diagram $h$ versus $J_{3}$ for the spin chain model of Eq. (\ref{parHamilt}), with $J=1$.}
\label{phasedig}
\end{figure}
%%%%%%%%%%%%%%%%%%%%%%%%%%%%%%%%%%%%%%%%%%%%%%%%
\begin{figure}[h!]
\centering
\includegraphics[width=0.47\textwidth]{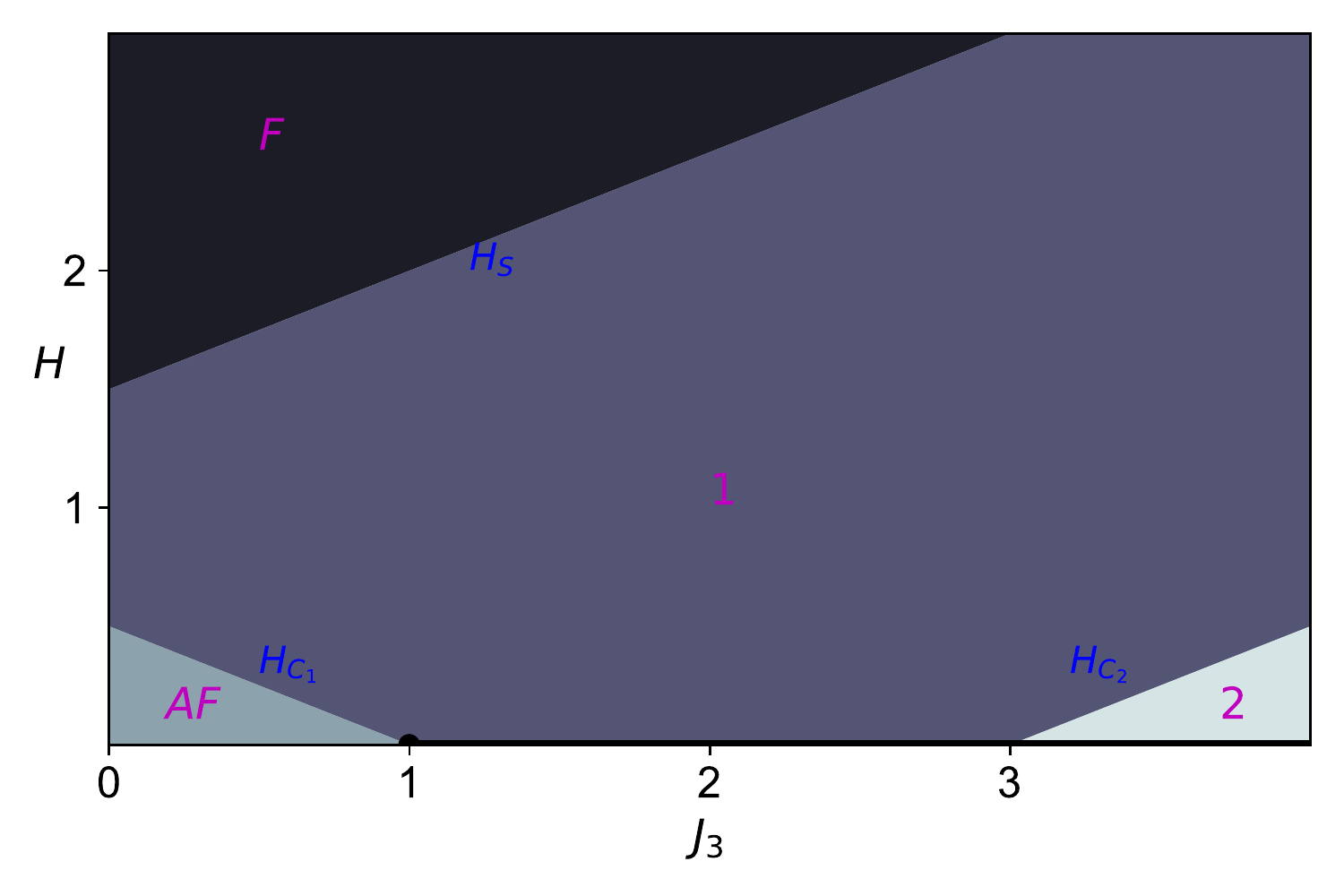}
\caption{The phase diagram $H$ vs. $J_{3}$ in the absence of four-spin interactions 
for $\mu_{1}=\mu_{2}=1$, $J_{1}=2$, $J_{2}=1$, $J_{13}=J_{23}=J_{3}\geq0$, and  $J_{14}=J_{24}=0$.}
\label{phasedig3spin}
\end{figure}
%%%%%%%%%%%%%%%%%%%%%%%%%%%%%%%%%%%%%%%%%%%%%%%% 
\subsection{The ground state and phase diagram}
        
Our model Eq. (\ref{parHamilt}), inspite of various simplifications, has a rich phase diagram which we show in 
Fig. (\ref{phasedig}) (here, we have set $J=1$). 
The ground state of any fermionic system corresponds to the situation where all possible states with positive 
energies are empty, and all states with negative energies are occupied. This ground state filling 
of two branches of Eq. (\ref{energy}) naturally depends on the values of the model parameters. 
Consider the non-negative values of the magnetic field parameter $h\geq0$ (from here and onwards, the 
magnetic field is specified by $h$ instead of $H$). There are four critical values of magnetic field, denoted 
by $h_{1,2,3}$ and $h_{13}$, and the explicit expressions of $h_{1,2,3}$ can be evaluated by putting ${\cal E}_{k,1,2}=0$ 
for $k=0$ and $\pi$, while for $h_{13}$, we need to maximise the solution of ${\cal E}_{k,1,2}=0$ with respect to $k$. 
We find 
\begin{eqnarray}
h_{1,2}&=&\frac{1}{3} \left(\pm\sqrt{12 J^2+J_{3}^2}+ 4 J_{3}\right), \notag\\
h_{3}&=&\frac{1}{3} \left(\sqrt{12 J^2+J_{3}^2}- 4J_{3}\right)~,\label{critical mag1}
\end{eqnarray}
and 
\begin{eqnarray}
h_{13}&=&\frac{\sqrt{24 J^2+J_{3}^2 \cos (2 k_{m})+J_{3}^2-12 \cos
(2 k_{m})+12}}{3 \sqrt{2}}+\notag\\
& &\frac{4}{3} J_{3} \cos (k_{m})~,\label{critical mag2}
\end{eqnarray}
where the expression of $k_{m}$ is lengthy, and is given in Appendix \ref{AppB}. First, consider the case with 
a small magnetic field, $0<h<h_{3}$, and values of three spin-exchange interactions, $0<J_{3}<2/\sqrt{5}$. 
Here, the critical line $h_{13}$ touches the line $h_{1}$ at $J_{3}=0.76$ for $J=1$, and $h_{2,3}$ 
intersect the $J_{3}$-axis at $J_{3}=(2\sqrt{J^{2}})/\sqrt{5}$. In the first branch, ${\cal E}_{k,1}$ is 
negative, and in the second branch, ${\cal E}_{k,2}$ is positive for any $k$. This domain is denoted by 
$\Romannum{1}$ in the phase diagram and is referred to as the ferrimagnetic phase, as can 
be gleaned from the ground state magnetisation per cell in the thermodynamic limit given by \cite{Zvyagin1,Zvyaginbook}
\begin{equation}
m_{\Romannum{1}}^{z}=1-\frac{1}{2\pi}\int_{-\pi}^{\pi}\left(\frac{\partial{\cal E}_{k,1}}{\partial h}\right)dk~.
\end{equation} 
In the region $h_{2,3}<h<h_{1}$, in the first branch, ${\cal E}_{k,1}<0$ for $|k|<k_{c_{1}}$ and ${\cal E}_{k,1}>0$ 
for $|k|>k_{c_{1}}$, while for the second branch, ${\cal E}_{k,2}>0$ for any $k\in(-\pi,\pi)$. This region is 
denoted as $\Romannum{2}$ in the phase diagram, and is a ferrimagnetic phase with gapless excitations. 
Similarly, the region between $h_{1}<h<h_{13}$ for $0<J_{3}<0.76$ denoted by $\Romannum{3}$ in the phase diagram 
is a ferrimagnetic phase in which the first branch ${\cal E}_{k,1}<0$, for 
$k\in(-k_{c_{1}},-k_{c_{2}})\cup(k_{c_{2}},k_{c_{1}})$, while in the second branch, 
${\cal E}_{k,2}>0$ for any $k\in(-\pi,\pi)$ with 
\begin{eqnarray}
k_{c_{1,2}}&=&\cos ^{-1}\Bigg(\frac{4 h J_{3}}{5
J_{3}^2+4}\mp\notag\\
& &\frac{\sqrt{h^2 J_{3}^2-12
h^2+20 J^2 J_{3}^2+16 J^2+20 J_{3}^2+16}}{5
J_{3}^2+4}\Bigg)~.\notag\\
\end{eqnarray}  
Consider now higher values of three spin-exchange interaction $J_{3}>2/\sqrt{5}$, and magnetic 
field in the range $0<h<h_{2}$. Here, for the first branch, ${\cal E}_{k,1}<0$ for $|k|<k_{c_{1}}$, and for the second branch 
${\cal E}_{k,2}<0$ for $|k|<k_{c_{2}}$. The system is in the  ferrimagnetic state with gapless excitations 
and is denoted by $\Romannum{4}$ in the phase diagram. Finally, for small values of $J_{3}<0.76$, 
both the branches are positive for $h>h_{13}$, while for higher values of $J_{3}>0.76$, both branches 
are positive for $h>h_{1}$. This region is denoted by $\Romannum{5}$ in the phase diagram, and the system 
will show ferromagnetic behaviour with gapped excitations. 

The ground state of the model of Eq. (\ref{parHamilt}) 
varies with parameters $h$ and $J_{3}$, and they are classified with distinct ranges of $k$ values in different 
regions of the phase diagram. The ground state of phase $\Romannum{1}$ can be described as a 
vacuum state with the first branch fully occupied and the second branch empty, i.e., 
$b_{k,1}\ket{\Psi}_{\Romannum{1}}=0$ and $b_{k,2}^{\dagger}\ket{\Psi}_{\Romannum{1}}=0$, 
where $\ket{\Psi}_{\Romannum{1}}=\prod\limits_{k}\ket{\Psi_{k,1,2}}_{\Romannum{1}}$, 
\begin{eqnarray}          
\ket{\Psi_{k,1,2}}_{\Romannum{1}}=-u_{k}\ket{1}_{k,1}\!\ket{0}_{k,2}+\frac{v_{k}(J\!-\!i\sin k)}
{\sqrt{J^{2}+\sin^{2}k}}\ket{0}_{k,1}\!\ket{1}_{k,2}~,\label{ground}\notag\\
\end{eqnarray}
with $k\in(-\pi,\pi)$. In region $\Romannum{2}$ of the phase diagram, $b_{k,1}\ket{\Psi}_{\Romannum{2}}=0$ 
for $|k|<k_{c_{1}}$ and $b_{k,2}^{\dagger}\ket{\Psi}_{\Romannum{2}}=0$ for $k\in(-\pi,\pi)$. This implies that 
the ground state in phase $\Romannum{2}$ is $\ket{\Psi}_{\Romannum{2}}=\prod\limits_{k}\ket{\Psi_{k,1,2}}_{\Romannum{1}}$ 
for $|k|<k_{c_{1}}$ and is trivial for other values of $k$ in this region. Similarly, in the region 
$\Romannum{3}$ and $\Romannum{4}$ of the phase diagram, the ground state is given by 
$\ket{\Psi}_{\Romannum{3},\Romannum{4}}=\prod\limits_{k}\ket{\Psi_{k,1,2}}_{\Romannum{1}}$, 
where $k\in(-k_{c_{1}},-k_{c_{2}})\cup(k_{c_{2}},k_{c_{1}})$. Note that the limits of $k$ in the ground states 
of the regions $\Romannum{3}$ and $\Romannum{4}$ are the same, but the values of $h$ and $J_{3}$ are different. 
The ground state of region $\Romannum{5}$ i.e., the ferromagnetic phase of the phase diagram is trivial since 
both the energy eigenvalue branches are positive for $k\in(-\pi,\pi)$, and is given by
$\ket{\Psi}_{\Romannum{5}}=\prod\limits_{k}\ket{1}_{k,1}\!\ket{1}_{k,2}$.
    
By the same method, for the sake of completeness, we also draw the phase diagram for the simpler case up to three spin interactions with 
parameters $\mu_{1}=\mu_{2}=1$, $J_{1}=2$, $J_{2}=1$, $J_{13}=J_{23}=J_{3}\geq0$, and  $J_{14}=J_{24}=0$. 
The ground-state phase diagram is shown in Fig. (\ref{phasedig3spin}), where F/AF denotes the 
ferromagnetic/antiferromagnetic phases, and 1 and 2 denote ferrimagnetic phases. 
The lines $H_{c_{1,2}}$ and $H_{s}$ are the lines of second-order QPT with critical values
\begin{eqnarray}
H_{c_{1}}&=&\frac{1-J_{3}}{2},\;\;H_{c_{2}}=\frac{J_{3}-3}{2},\;\;H_{s}=\frac{J_{3}+3}{2}~.
\end{eqnarray}
Also, the line $H=0$, $J_{3}>1$ is the line of first order QPT, i.e., the spontaneous magnetisation 
changes from zero to a finite value.

\section{\label{sec:level1}The QIM and complexity}
The ground state wavefunction Eq. (\ref{ground}) is defined on the parameter space $h$ and $J_{3}$ 
for a fixed value of $J=1$. The QIM measures the distance between two quantum states $\ket{\Psi(\vec{\lambda})}$ 
and $\ket{\Psi(\vec{\lambda}+\vec{d\lambda})}$, where $\lambda=\{h, J_{3}\}$ and $\vec{d\lambda}$ is the infinitesimal 
separation in parameter space. The distance can be expressed as:
\begin{eqnarray}
d\tau^{2}&=&\sum\limits_{i,j}g_{ij}d\lambda^{i}d\lambda^{j}~,
\end{eqnarray}
with $i=1,2,....,m$, and $m$ being the dimensions of parameter space, which in our case is 2. 
We now find the QIM in the regions of the ground state phase diagram. 

For small values of $J_{3}$ in region $\Romannum{1}$, we have calculated the QIM numerically in the thermodynamic limit, 
and found that the numerical results are in good agreement with the analytical expressions. In all other regions, 
we only rely on the numerical results due to the complicated expressions of $k_{c_{1,2}}$. 
We now turn to the QIM of region $\Romannum{1}$, which takes the following form:
\begin{eqnarray}
g_{hh}&=&\frac{1}{4}\!\sum\limits_{k}\left(\frac{\partial\theta_{k}}{\partial h}\right)^{2}~,~~
g_{J_{3}J_{3}}=\frac{1}{4}\sum\limits_{k}\left(\frac{\partial\theta_{k}}{\partial J_{3}}\right)^{2},\notag\\
g_{hJ_{3}}&=&g_{J_{3}h}=\frac{1}{4}\sum\limits_{k}\left(\frac{\partial\theta_{k}}{\partial h}\right)
\left(\frac{\partial\theta_{k}}{\partial J_{3}}\right)~,\label{metric}
\end{eqnarray} 
%%%%%%%%%%%%%%%%%%%%%%%%%%%%%%%%%%%%%%%%%%%%%%%%
\begin{figure}[h!]
\centering
\includegraphics[width=0.35\textwidth]{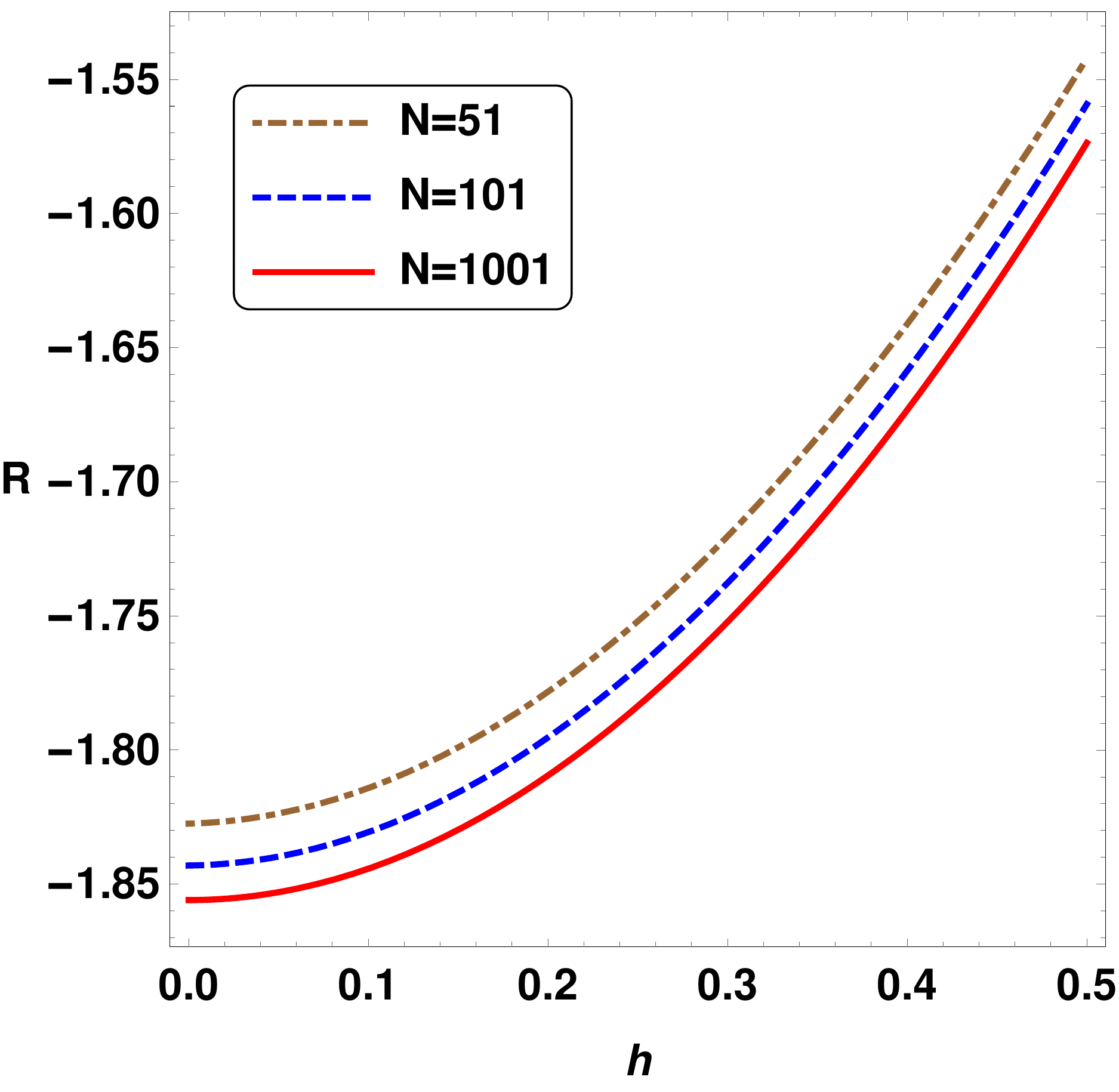}
\caption{Ricci scalar $R$ as a function of $h$ of region $\Romannum{1}$, for $J_{3}=0.5$, with the 
system sizes $N=51$ (dot-dashed brown), $N=101$ (dashed blue), and $N=1001$ (solid red).}
\label{Ricci}
\end{figure}
%%%%%%%%%%%%%%%%%%%%%%%%%%%%%%%%%%%%%%%%%%%%%%%%
where $k\in(-\pi,\pi)$. In the thermodynamic limit, the procedure to evaluate these 
is simple : the summations in the above metric components are converted into integrals, i.e., we replace 
$\sum\limits_{k}=\frac{N}{2\pi}\int_{-\pi}^{\pi}dk$, and then do a standard computation of residues in the complex plane. 
The analytic expressions of the metric components are too lengthy to be presented here. However, we have expanded 
these metric components by assuming a small value of $J_{3}$ and then analysed the Ricci scalar, $R$. Even then 
the analytic expression for $R$ is cumbersome, and will not be presented here. Numerical values of $R$ are shown in 
Fig. (\ref{Ricci}). Importantly, there is no divergence on the critical line $h_{3}$. 
 %%%%%%%%%%%%%%%%%%%%%%%%%%%%%%%%%%%%%%%%%%%%%%%%
\begin{figure}[h!]
\centering
\includegraphics[width=0.35\textwidth]{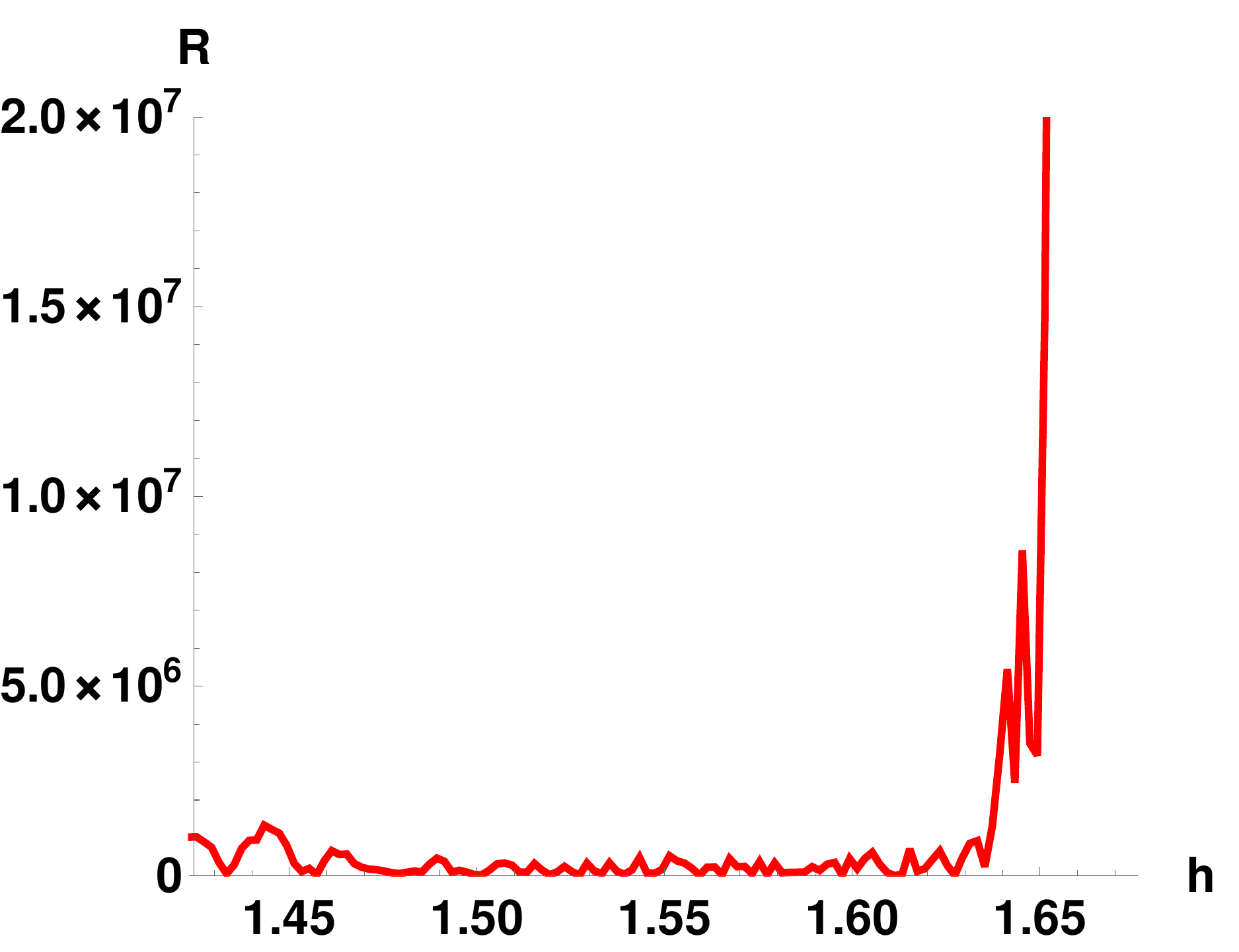}
\caption{Ricci scalar $R$ as a function of $h$ of region $\Romannum{3}$, for $J_{3}=0.2$, with system size $N=101$.}
\label{Riccireg3}
\end{figure}
%%%%%%%%%%%%%%%%%%%%%%%%%%%%%%%%%%%%%%%%%%%%%%%% 

In region $\Romannum{2}$, the components of the QIM take the same form as given in Eq. (\ref{metric}), but here 
$k\in(-k_{c_{1}}, k_{c_{1}})$. We have computed the Ricci Scalar numerically, which shows oscillations 
in this region and is divergent on approaching the critical line $h_{1}$. These oscillations in $R$ 
are not particularly interesting, as these are due to 
the discontinuity in the metric tensor components, since $k$ is not summed over the entire range $(-\pi,\pi)$ 
in the metric tensor : the summation here runs for $k$ in the range $(-\frac{2\pi\lambda_{c_{1}}}{N},
\frac{2\pi\lambda_{c_{1}}}{N})$ where $\lambda_{c_{1}}=\frac{Nk_{c_{1}}}{2\pi}$. The integer values of 
$\lambda_{c_{1}}$ are different for distinct values of $h$ and $J_{3}$. For example, if $(J_{3}, h)=(2,1.2)$, 
then $\lambda_{c_{1}}=14.05$, and the summation runs over 0 to 14, while for $(J_{3}, h)=(2,1.4)$, 
$\lambda_{c_{1}}=13.45$, and summation runs over 0 to 13. This creates discontinuities in the nature of the metric tensor, 
which is reflected as oscillatory nature in the Ricci scalar plots.

Similarly, in region $\Romannum{3}$, the summation in QIM tensor in Eq. (\ref{metric}) runs 
over $(-k_{c_{1}},-k_{c_{2}})\cup(k_{c_{2}},k_{c_{1}})$. In this region also, the Ricci scalar shows an apparently 
oscillatory nature, and has a clear divergence on approaching the critical line $h_{13}$. The behaviour of $R$ as a 
function of $h$ is shown in Fig. (\ref{Riccireg3}). In the same way, we have computed $R$ in region $\Romannum{4}$. 
Here, it turns out that there is no divergence on the critical line $h_{2}$, which indicates that the geometry of 
the ground state is regular everywhere except on the critical lines $h_{13}$, and $h_{1}$ for $J_{3}>0.76$, i.e., transition 
lines which separate the ferrimagnetic from ferromagnetic phases.  

\subsection{The geodesics and FSC}

Now we numerically compute the geodesics and FSC in the ground state phase diagram. The method is standard and has
been described in details for spin systems in \cite{tapo}. Namely, to obtain the geodesics, 
we parametrise $h$ and $J_{3}$ as a function of an affine parameter $\tau$, and then we solve the 
two coupled second order differential equations given from
\begin{equation}
\frac{d^{2}x^{i}}{d\tau^{2}}+\Gamma^{i}_{jl} \frac{dx^{j}}{d\tau} \frac{dx^{l}}{d\tau}=0~,
\end{equation}
%%%%%%%%%%%%%%%%%%%%%%%%%%%%%%%%%%%%%%%%%%%%%%%%
\begin{figure}[h!]
\centering
\includegraphics[width=0.35\textwidth]{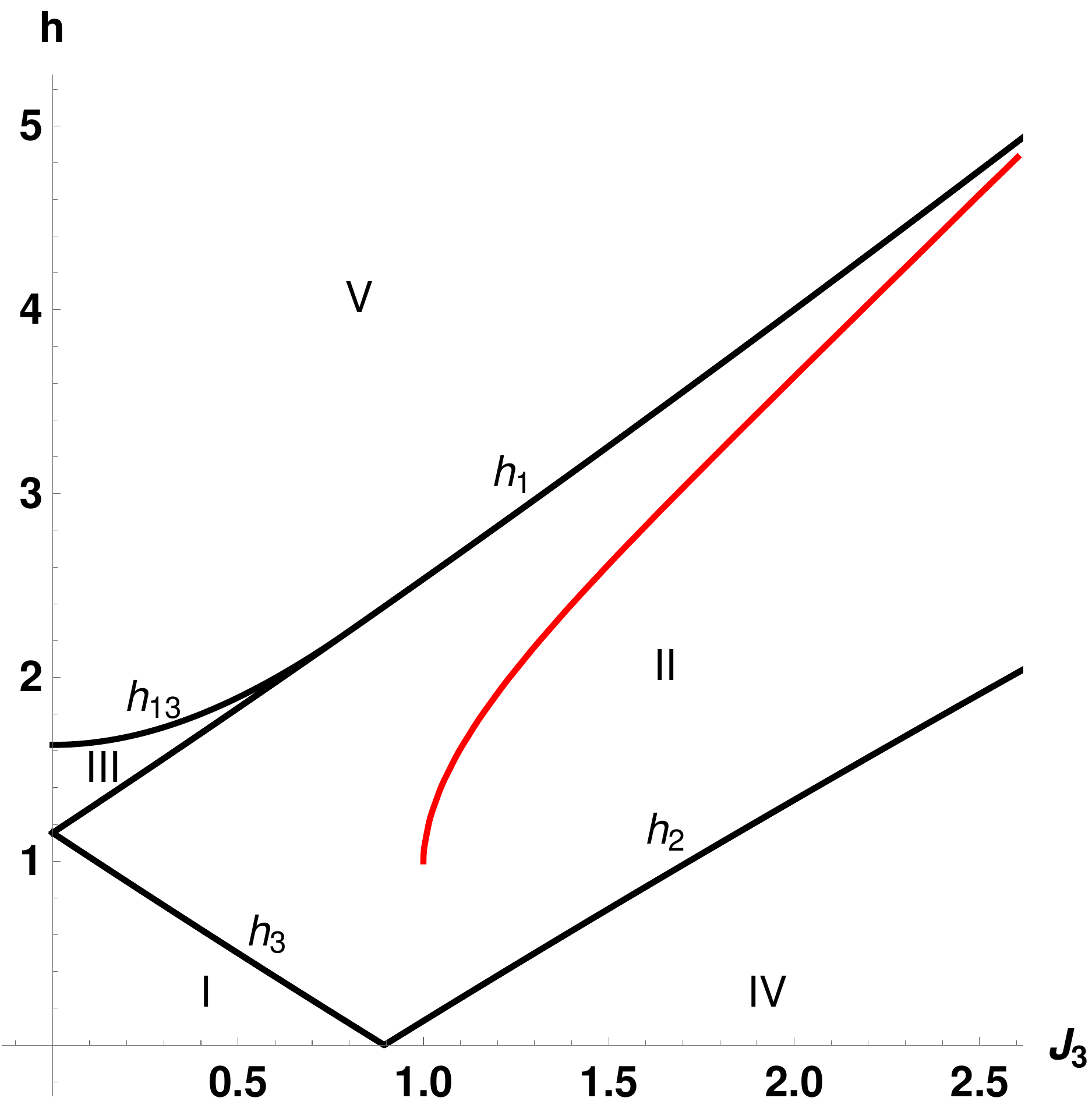}
\caption{Geodesic on $h-J_{3}$ plane in region $\Romannum{2}$ for system size $N=51$ is shown in solid red, 
and critical lines are in solid black.}
\label{geodesic}
\end{figure}
%%%%%%%%%%%%%%%%%%%%%%%%%%%%%%%%%%%%%%%%%%%%%%%% 
where $x^{i}=\{h(\tau), J_{3}(\tau)\}$, and $\Gamma^{i}_{jl}$ are Christoffel symbols. 
We solve these two equations numerically in each region of the phase diagram by specifying the values of 
four initial conditions $h(\tau=0)$, $J_{3}(\tau=0)$, $\dot{h}(\tau=0)$, $\dot{J_{3}}(\tau=0)$, and one normalisation 
condition $g_{ij}\dot{x}^{i}\dot{x}^{j}=1$, with dot implies a derivative with respect to $\tau$. This 
then determines $h(\tau)$ and $J_{3}(\tau)$. In regions $\Romannum{1}$ and $\Romannum{4}$, the geodesics 
do not show any special behaviour since the metric and the Ricci scalar are regular throughout these regions. 
We have inverted the solution of geodesics to find the FSC, i.e., $\tau(h)$, using a standard root finding procedure
in Mathematica. These do not have any special behaviour up to the phase boundary $h_{2,3}$, reflecting that the 
ground state manifold is regular on the critical line $h_{2,3}$. 
%%%%%%%%%%%%%%%%%%%%%%%%%%%%%%%%%%%%%%%%%%%%%%%%
\begin{figure}[h!]
\centering
\includegraphics[width=0.35\textwidth]{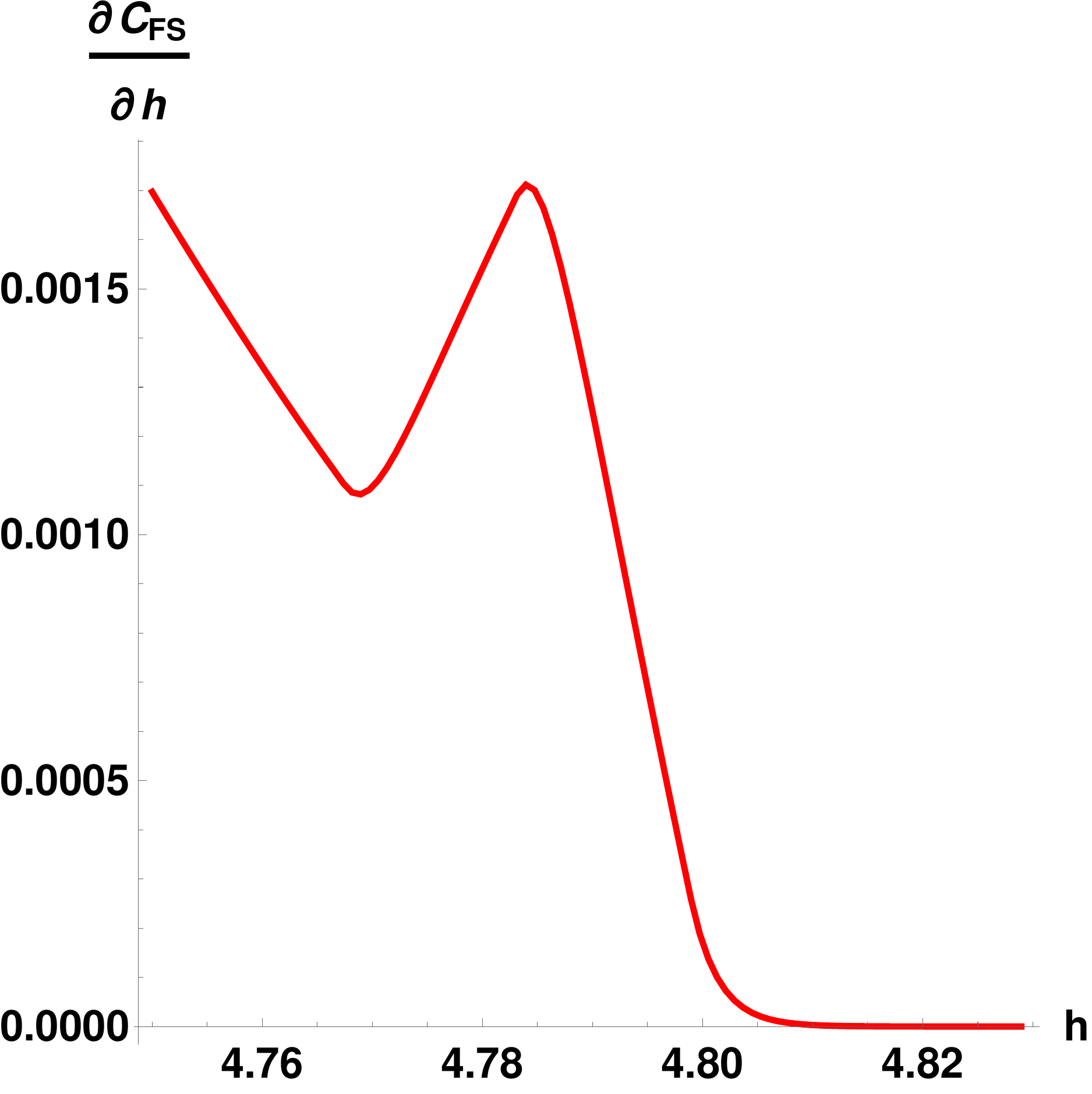}
\caption{The derivative $\partial\mathcal{C}_{FS}/\partial h$ as a function of $h$ of region 
$\Romannum{2}$, for system size $N=51$.}
\label{fscder}
\end{figure}
%%%%%%%%%%%%%%%%%%%%%%%%%%%%%%%%%%%%%%%%%%%%%%%% 

In region $\Romannum{2}$, the behavior of a typical geodesic is shown in Fig. (\ref{geodesic}). 
In our numerical analysis, we have chosen the initial conditions $h(\tau=0)=1$, $J_{3}(\tau=0)=1$, 
$\dot{J_{3}}(\tau=0)=0.04$, and $\dot{h}(\tau=0)=6.78$ is determined from the normalisation condition. 
The geodesics that start from any coordinate $(J_{3}, h)$ where $J_{3}<0.76$ and $h_{3}<h<h_{1}$ approach the 
phase boundary $h_{1}$ but do not show any interesting behaviour. On the other hand, the geodesics that start 
from the $J_{3}>0.76$ and $h>h_{2}$ keeps approaching the critical line $h_{1}$, 
attains a minimum distance with the critical line, but will never cross the phase boundary, as shown in 
Fig. (\ref{geodesic}). The FSC, and its derivative are numerically evaluated, and the derivative, 
$\partial\mathcal{C}_{FS}/\partial h$ is plotted in Fig. (\ref{fscder}). The derivative of FSC approaches zero 
when the distance between the geodesic and phase boundary line is minimal. This can be explained by 
a simple relation that we have obtained previously in \cite{tapo}, namely,
\begin{equation}
\frac{\partial\tau}{\partial h}=\frac{\partial\mathcal{C}_{FS}}{\partial h}=\left(g_{hh}\right)^{\frac{1}{2}}~.
\end{equation}
Here, $g_{hh}\to 0$ at $h\to h_{1}$ for $J_{3}>0.76$ and $J=1$ is fixed. 
The behaviour of geodesics in region $\Romannum{3}$ is almost similar to region $\Romannum{2}$ for $J_{3}>0.76$. 
The geodesics attain a minimum distance with the critical line $h_{13}$, but after that, it starts approaching 
the line $h_{1}$ for $J_{3}<0.76$. The derivative of FSC with respect to $h$, 
$\frac{\partial\mathcal{C}_{FS}}{\partial h}\to 0$ at $h \to h_{13}$. 
%%%%%%%%%%%%%%%%%%%%%%%%%%%%%%%%%%%%%%%%%%%%%%%%
\begin{figure}[h!]
\centering
\includegraphics[width=0.35\textwidth]{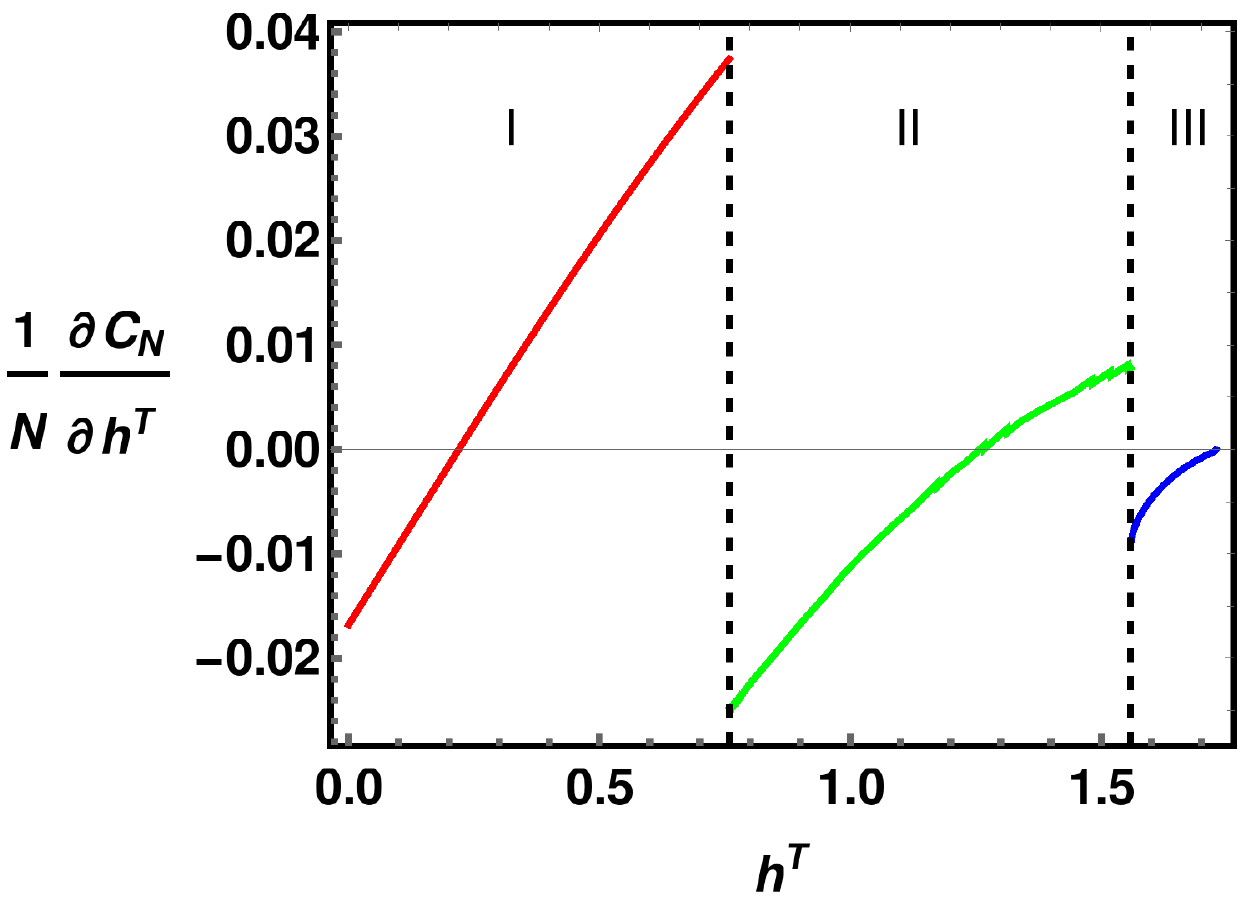}
\caption{The derivative of NC with respect to $h^{T}$ versus $h^{T}$ for $J=1$, 
$N=101$, $J_{3}^{R}=0.1$, and $J_{3}^{T}=0.3$. Here, $h^{R}=0.2, 1.05, 1.06$ in regions
I, II, and III, respectively.}
\label{derNc}
\end{figure}
%%%%%%%%%%%%%%%%%%%%%%%%%%%%%%%%%%%%%%%%%%%%%%%%

\subsection{The Nielsen Complexity}
For a fixed value of $J=1$, the ground state of the model can be expressed in the form:
\begin{eqnarray}          
\ket{\Psi}=\prod_{k}\left[-u_{k}\ket{1}_{k,1}\!\ket{0}_{k,2}+v_{k}e^{-i\phi_{k}}\ket{0}_{k,1}\!\ket{1}_{k,2}\right]~,\notag\\
\end{eqnarray}
where $e^{-i\phi_{k}}=(1-i\sin k)/\sqrt{1+\sin^{2}k}$. By following \cite{Liu}, the NC of our model 
can be computed straightforwardly, i.e., $\mathcal{C}_{N}=\sum_{k}|\Delta\theta_{k}|^{2}$, 
where $\Delta\theta_{k}=(\theta_{k}^{T}-\theta_{k}^{R})/2$, and 
\begin{equation}
\cos \theta_{k}^{R,T}=\frac{\frac{h^{R,T}}{2}-J_{3}^{R,T}\cos k}
{\sqrt{\left(\frac{h^{R,T}}{2}-J_{3}^{R,T}\cos k\right)^{2}+1+\sin^{2}k}}~.
\end{equation}  
Hereafter, the superscripts $R$ and $T$ on a variable are used to denote the reference and the target values 
respectively, of that variable. Note that 
unlike the Kitaev model \cite{Liu} or the XY spin chain \cite{tapo}, 
the derivative of the NC does not have any divergences on any of the critical 
lines $h_{1,2,3,13}^{T}$. This derivative of NC with respect to $h^{T}$ is plotted in Fig. (\ref{derNc}) 
as a function of $h^{T}$. It can be seen that the derivative approaches zero on the critical line $h_{13}^{T}$, 
in region $\Romannum{3}$ of the phase diagram. This behaviour can be explained using the relation \cite{tapo}
\begin{equation}
\frac{\partial\mathcal{C}_{N}}{\partial h^{T}}\sim\int\sum_{k}\left(\frac{\partial\theta_{k}^{T}}
{\partial h^{T}}\right)^{2}dh^{T}=\int g_{h^{T}h^{T}}dh^{T}~.
\end{equation}
Since $g_{h^{T}h^{T}}\to0$ on $h^{T}\to h_{13}^{T}$ , so the derivative of NC goes to zero on this critical line. 
A similar behaviour is observed as $h^{T}\to h_{1}^{T}$ for $J_{3}^{T}>0.76$ in region $\Romannum{2}$.

\section{\label{LE}The Time-dependent Nielsen Complexity and Loschmidt Echo}

We now consider the model Eq. (\ref{parHamilt}) with a quantum quench, assuming the sudden quench approximation. 
Following the work of \cite{ZanLE}, we consider our model
coupled to a two-level central spin-1/2 system. Here, we will take the interaction Hamiltonian to be ${\cal H }_I$, 
so that the total Hamiltonian is ${\cal H }_F={\cal H }+{\cal H }_I$,
with ${\cal H }$ given in Eq. (\ref{parHamilt}), and 
\begin{eqnarray}
{\cal H }_I &=&-\frac{\delta}{2}\ket{e}\bra{e}\sum\limits_{n}\left(3S_{n,1} ^{z}+S_{n,2}^{z}\right)~,
\label{singlequench}
\end{eqnarray}
where $\delta$ is the interaction strength, and $\ket{e}$ is the excited state of the two-level system. 
By a standard method, we write the ground state $\ket{\Psi}_{h,J_{3}}$ 
of ${\cal H }$ in terms of that of ${\cal H }_F$, labeled $\ket{\Psi_{k,1,2}}_{h+\delta,J_{3}}$ 
for the $k$-th Fourier mode. This gives
\begin{equation}
\ket{\Psi}_{h,J_{3}}=\prod_{k}\left[\cos\Omega_k +\sin\Omega_k\chi_{k,1}^{\dagger}
\chi_{k,2}\right]\ket{\Psi_{k,1,2}}_{h+\delta,J_{3}}~,
\end{equation}
where we have defined 
\begin{eqnarray}
\Omega_k &=& \frac{1}{2}\left[\theta_k(h,J_{3}) - \theta_k(h+\delta,J_{3})\right],\nonumber\\
\chi_{k,1} &=&e^{-i\phi_{k}}v_{k}(h+\delta,J_{3})d_{k,1}+u_{k}(h+\delta,J_{3})d_{k,2},\nonumber\\
\chi_{k,2} &=&-e^{-i\phi_{k}}u_{k}(h+\delta,J_{3})d_{k,1}+v_{k}(h+\delta,J_{3})d_{k,2}~,\notag\\
\end{eqnarray}
with the operators $d_{k,1}$ and $d_{k,2}$ being the Fourier operators, and $\theta_{k}$, $u_{k}$, 
and $v_{k}$ are defined in Eq. (\ref{bogoliubov}) with the arguments appropriately shifted. 

To compute the complexity, the reference and the target states are chosen to be $\ket{\Psi}_{h,J_{3}}$
and $\ket{\Psi_{e}(t)}$ respectively, where, $\ket{\Psi_{e}(t)}=e^{-i{\cal H }_{F}(h+\delta,J_{3})t}\ket{\Psi}_{h,J_{3}}$, i.e.,
\begin{eqnarray}
\ket{\Psi_{e}(t)}&=&\prod\limits_{k}\bigg[e^{-i{\cal E}_{k,2}(h+\delta,J_{3})t}\cos(\Omega_{k})
+e^{-i{\cal E}_{k,1}(h+\delta,J_{3})t}\times\notag\\
& &\sin(\Omega_{k})
\chi_{k,1}^{\dagger}\chi_{k,2}\bigg]\ket{\Psi_{k,1,2}}_{h+\delta,J_{3}}~.
\label{wavefn}
\end{eqnarray}
The computation of the NC is now standard (see, e.g., \cite{Liu}). The final result after the quench is  
\begin{equation}
{\mathcal C}_{N}(t)\equiv \sum\limits_{k}{\mathcal C}_{Nk}=\sum\limits_{k}\Phi_{k}^{2}(h+\delta,J_{3},t)~,
\label{compt}
\end{equation}
where \begin{equation}
\Phi_{k}=\arccos\left(\sqrt{1-\sin^{2}(2\Omega_{k})\sin^{2}(\Lambda_{k}(h+\delta,J_{3})t)}\right)~,
\label{Phik}
\end{equation}
with the single particle excitations 
$\Lambda_{k}(h+\delta,J_{3}) = \sqrt{\left(\frac{h+\delta}{2}-J_{3}\cos k\right)^{2}+1+\sin^{2}k}$. 
A particularly interesting quantity in a system after quench is the Loschmidt Echo. 
The system is initially prepared in the ground state $\ket{\Psi}_{h,J_{3}}$ of the Hamiltonian ${\cal H }$, 
and it is then quenched to evolve according to the Hamiltonian ${\cal H }_{F}$. The LE is defined as
\begin{eqnarray}
{\mathcal L}&=&\big|\,_{h,J_{3}}\!\!\bra{\Psi}e^{-i{\cal H }_{F}(h+\delta,J_{3})t}\ket{\Psi}_{h,J_{3}}\big|^2,\notag\\
&=&\prod_{k}\big[1-\sin^{2}(2\Omega_{k})\sin^{2}(\Lambda_{k}(h+\delta,J_{3})t)\big]~,
\end{eqnarray}
\begin{figure}[h!]
\centering
\includegraphics[width=0.45\textwidth, height=0.4\textwidth]{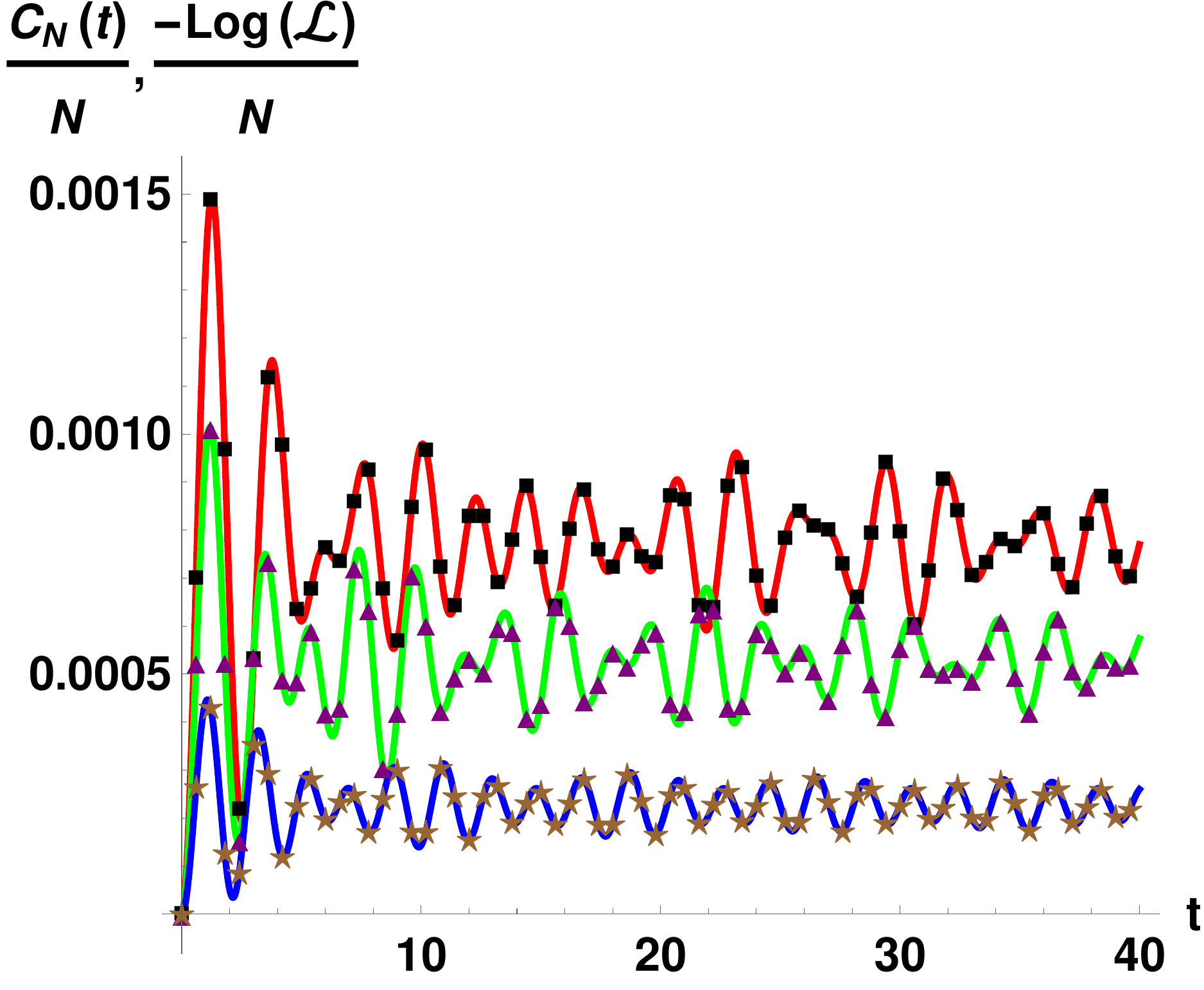}
\caption{Comparison between time-dependent NC (solid lines) and LE (scattered plots). The red 
(black squares), green (purple triangles), and blue (brown stars) correspond to $h=0.5$, 1, and 1.4, 
respectively, for fixed values of $N=101$, $\delta=0.1$, $J_{3}=0.2$, and $J=1$.}
\label{LeNC}
\end{figure}
As discussed in \cite{tapo1}, the relation between LE and NC for small times, ${\mathcal L}\simeq e^{-{\mathcal C}_{N}}$ 
is also valid for our spin chain model. In fact, Fig. (\ref{LeNC}) shows that this relation continues to hold 
for large times as well. We have plotted the dynamical behavior of the time-dependent NC per system size, 
${\mathcal C}_{N}(t)/N$ along with negative of logarithm of LE per system size, $-\log({\mathcal L})/N$ for $N=101$, 
$\delta=0.1$, $J_{3}=0.2$, and $J=1$. The solid red (scattered black squares), solid green (scattered purple triangles), 
and solid blue (scattered brown stars) shows the numerical results for ${\mathcal C}_{N}(t)/N$ ($-\log({\mathcal L})/N$) 
corresponding to $h=0.5$, 1, and 1.4, for regions $\Romannum{1}$, $\Romannum{2}$, and $\Romannum{3}$ 
of the phase diagrams respectively. 

As anticipated, the time-dependent NC  first increases linearly and then starts oscillating \cite{Liu, tapo1}. 
However, in our model, the oscillations do not die out to approach a time-independent value, contrary to what is observed 
in those works. This oscillatory behaviour is observed here in all the regions of the phase diagram for any 
finite or large times and is well explained by the results in \cite{tapo1}. 
That is, the time-dependent NC and LE are oscillatory functions of $k$ due 
to the nature of $\Omega_{k}$ and $\Lambda_{k}$. The oscillations die out rapidly in the 
cases considered in \cite{Liu, tapo1} because a large number 
of Fourier modes start contributing with the increase in time, and they interfere destructively. But in our 
four-spin interaction model, the Fourier modes are asymmetrical in $k$-space. As a result, even if there are a large number 
of maxima and minima interfering with one another with increase in time, some parts of the contributing modes 
still cannot interfere in a completely destructive manner. Hence, temporal oscillations are always present in 
the system. Secondly, the relation ${\mathcal L}= e^{-{\mathcal C}_{N}}$ is valid for all times and in all the 
regions of the phase diagram. This is because the term $\sin^{2}(2\Omega_{k})<<1$ everywhere in the phase diagram 
and even on the critical lines $h_{1,2,3,13}$, unlike in \cite{tapo1}, where the relation is found to be invalid for 
the critical lines or when the quench protocol is on the critical lines.

We have also performed the analysis regarding the role of time-dependent NC and LE to identify the critical 
lines. For our model of Eq. (\ref{parHamilt}), we find that there is no 
non-analyticity or discontinuity in these quantities, whenever the phase change occurs. 
This reflects the fact that, like the NC, these two dynamical quantities cannot signal any particular characteristics 
of critical behaviour of the model.

\section{Multiple quench scenarios}
\label{multi}

In this section, we consider multiple quench scenarios in our model, extending the 
analysis of section \ref{LE}. 
We will consider here a series of four sudden quenches in the model. 
The multiple quench protocol is as follows. 
At time $t=0$, the external magnetic field $h$ is quenched from
$h_{0}$ to $h_{0}+\delta$ for a time $t_{1}=15$s, where $\delta$ is the interaction
strength. At time $t_{1}$, the external magnetic field after quench i.e.,
$h_{0}+\delta$ is changed back to $h_{0}$ for a time $t_{2}=15$s, which is defined
as no quench in the system. This sequence has been repeated four times. 
The multiple quench protocol involves the interaction
Hamiltonian to be ${\cal H }_I^{multi}$, so that similar to Eq. (\ref{singlequench}), the total Hamiltonian is ${\cal H
}_F^{multi}={\cal H }+{\cal H }_I^{multi}$, with
\begin{eqnarray}
{\cal H }_I^{multi} &=&-\frac{\delta}{2}\ket{e}\bra{e}\sum\limits_{n}\left(3S_{n,1}
^{z}+S_{n,2}^{z}\right)F(t)~,
\end{eqnarray}
where we define
\begin{eqnarray}
F(t)&=&\left\{
\begin{array}{ll}
\Theta\left(\sin\left(\frac{\pi t}{T}\right)\right),&\,0<t<90\\
1,&\,t>90
\end{array}
\right.
\notag.
\end{eqnarray}
Here, the quench time is taken as $T=15$s, $\Theta$ is the Heaviside function, and $\ket{e}$ is
the excited state of the two-level system as discussed in the single quench
scenario. The ground state of the system and the time-dependent NC after time $t=T$
have been discussed in details in the single quench case. The ground state after time
$t=2T$ i.e., when $h_{0}+\delta$ is changed back to $h_{0}$ is
\begin{eqnarray}
\ket{\Psi_{e}(t_{1},t_{2})}&=&\prod\limits_{k}\bigg[e^{-i{\cal
E}_{k,2}(h,J_{3})t_{2}}Y_{k}
+e^{-i{\cal E}_{k,1}(h,J_{3})t_{2}}\times\notag\\
& & W_{k}
R_{k,1}^{\dagger}R_{k,2}\bigg]\ket{\Psi_{k,1,2}}_{h,J_{3}}~,
\end{eqnarray}
where we have defined 
\begin{eqnarray}
Y_{k} &=&e^{-i{\cal E}_{k,2}(h+\delta,J_{3})t_{1}}\cos^{2}\Omega_{k}+e^{-i{\cal
E}_{k,1}(h+\delta,J_{3})t_{1}}\sin^{2}\Omega_{k},\nonumber\\
W_{k} &=&\left(-e^{-i{\cal E}_{k,2}(h+\delta,J_{3})t_{1}}+e^{-i{\cal
E}_{k,1}(h+\delta,J_{3})t_{1}}\right)\sin\Omega_{k}\cos\Omega_{k},\nonumber\\
R_{k,1} &=&e^{-i\phi_{k}}v_{k}(h,J_{3})d_{k,1}+u_{k}(h,J_{3})d_{k,2},\nonumber\\
R_{k,2} &=&-e^{-i\phi_{k}}u_{k}(h,J_{3})d_{k,1}+v_{k}(h,J_{3})d_{k,2}~,\notag\\
\end{eqnarray}
with the operators $d_{k,1}$ and $d_{k,2}$ being the Fourier operators, and
$\theta_{k}$, $u_{k}$, and $v_{k}$ are defined earlier with the arguments
appropriately shifted. 

To compute the complexity, the reference and the target states are chosen to be
$\ket{\Psi}_{h,J_{3}}$
and $\ket{\Psi_{e}(t_{1},t_{2})}$ respectively. The final result of NC after $t=2T$ is 
\begin{equation}
{\mathcal C}_{N}(t_{1},t_{2})\equiv \sum\limits_{k}{\mathcal
C}_{Nk}(t_{1},t_{2})=\sum\limits_{k}\left(\arccos\left(\sqrt{Y_{k}^{\dagger}Y_{k}}
\right)\right)^{2}~.
\end{equation}
%%%%%%%%%%%%%%%%%%%%%%%%%%%%%%%%%%%%%%%%%%%%%%%%
\begin{figure}[h!]
\centering
\includegraphics[width=0.35\textwidth]{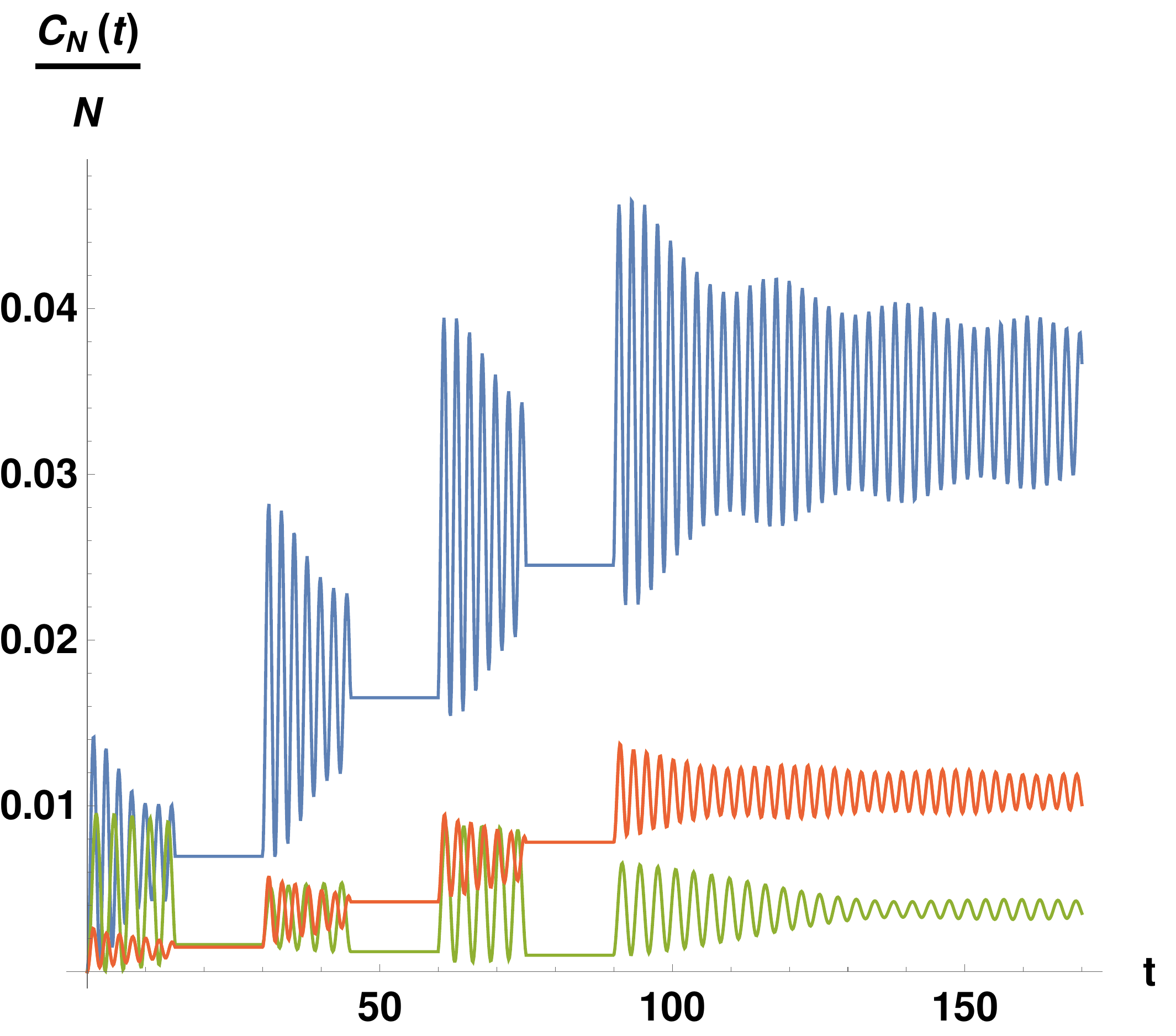}
\caption{The NC as a function of time is plotted for multiple (four) sudden quenches
with system size $N=501$, $J_{3}=1.5$, and $T=15$s. The blue, orange and green
curves corresponds to $h_{0}=$1, 1, 3, and $\delta=$-0.4, -0.2, 0.2, respectively.}
\label{multiquench}
\end{figure}
%%%%%%%%%%%%%%%%%%%%%%%%%%%%%%%%%%%%%%%%%%%%%%%%  
The second and subsequent quenches become difficult to handle analytically, and 
we have performed numerical analysis for the same. The
behaviour of the NC with time after four quenches is shown in Fig.
(\ref{multiquench}), for system size $N=501$, $J_{3}=1.5$, and $T=15$s. The time
evolution of the NC exhibits an oscillating behaviour whenever the system is quenched,
and the magnitude of oscillations decreases with time. Also, the magnitude of NC is
found to be constant whenever $h_{0}+\delta$ is changed back to $h_{0}$. In Fig.
(\ref{multiquench}), the blue curve corresponds to $h_{0}=1$ and $\delta=-0.4$; the
orange curve corresponds to $h_{0}=1$ and $\delta=-0.2$, and the green curve
corresponds to $h_{0}=3$ and $\delta=0.2$. Note that the green curve is plotted by
choosing parameter values such that the system comes on the critical line $h_{1}$
after the quench. As concluded in the single quench case, the NC in multiple quench
scenarios also does not indicate any critical behaviour of the system.
%%%%%%%%%%%%%%%%%%%%%%%%%%%%%%%%%%%%%%%%%%%%%%%%
\begin{figure}[h!]
\centering
\includegraphics[width=0.35\textwidth]{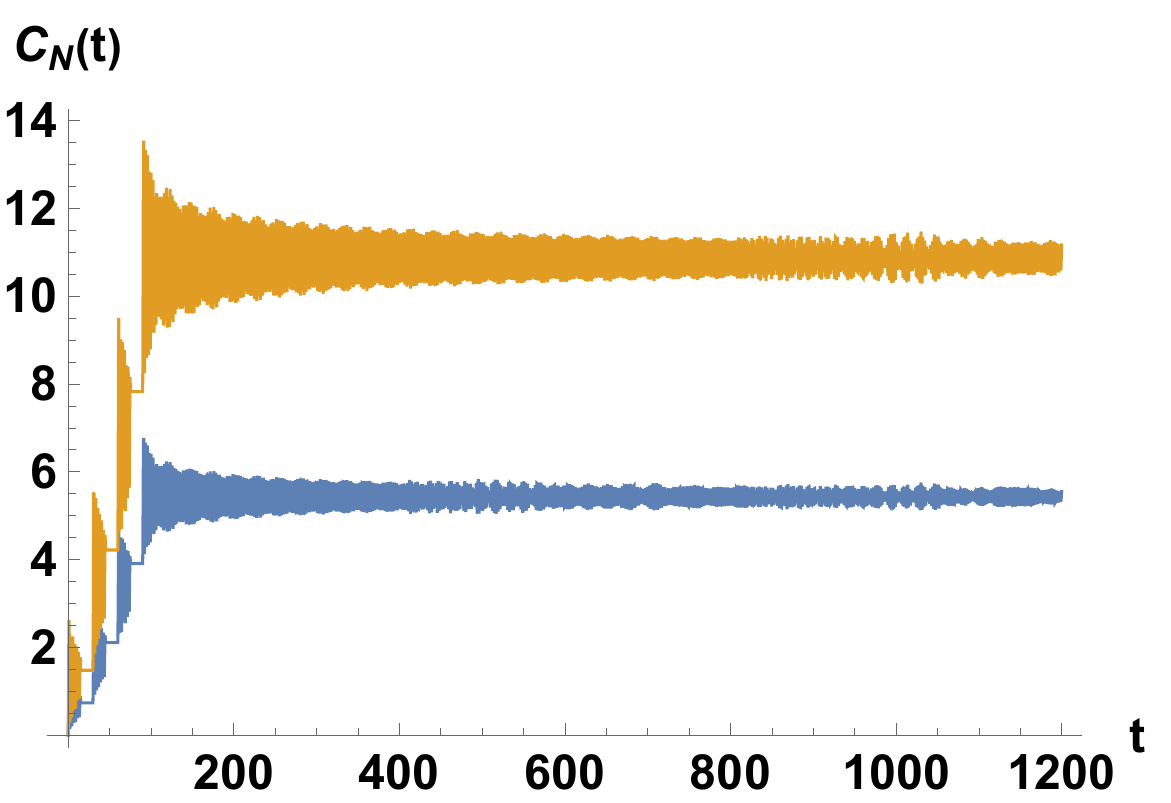}
\caption{The NC as a function of time is plotted for multiple (four) sudden quenches
for $J_{3}=1.5$, and $T=15$s with 
$h_{0}=1$, and $\delta=-0.2$. The blue and orange lines correspond to system sizes
$N=501$ and $1001$, respectively.}
\label{multiquench1}
\end{figure}
%%%%%%%%%%%%%%%%%%%%%%%%%%%%%%%%%%%%%%%%%%%%%%%%  
In Fig. (\ref{multiquench1}), we plot the NC at large times for $h_{0}=1$, and $\delta=-0.2$,
for system size $N=501$ (blue) and $N=1001$ (orange). We observe here that for late times
and for large system sizes, in our four spin interaction model, oscillations are regular, and their 
maximum amplitude decreases initially with time and later acquires nearly a constant value. 
These results are similar to what was observed in \cite{DG,tapo5} for integrable models.

We have also performed the same analysis of multiple quenches in the transverse
spin-1/2 XY model \cite{zanardi}, where the quench parameters are the anisotropic coupling and
transverse magnetic field. When we quench the value of the magnetic field, the behaviour of the NC
is similar to the one we encountered before -- like in the four-spin exchange interaction model, the
NC here first increases linearly and then starts oscillating over time. The
oscillations are more rapid when the system is quenched on one of the critical
lines. However, no other signal of QPT is observed from the complexity versus time
curve, even for the transverse spin-1/2 XY model.

\section{The DMRG and Entanglement Entropy}
\label{entanglement}

We now compute the entanglement entropy of the system using DMRG methods, which we first recall. 
The DMRG is a variational method that uses the matrix product states (MPS) and matrix product operator (MPO) representations 
to find the lowest energy state of a quantum many-body system. An MPS represents a quantum state with a product 
of matrices. The Hilbert space of the one-dimensional spin-1/2 chain has $2^{N}$ degrees of freedom for 
$N$ spins. The MPS can be manipulated to form a subspace of the larger $2^{N}$ dimensional Hilbert 
space which captures the essential physics. The search for the ground state using DMRG is 
an iterative solution of the Schrodinger equation ${\cal H }\ket{\psi}=E\ket{\psi}$. In the first step, 
the Hamiltonian of the system has to be constructed as an MPO. To solve the Schrodinger equation for $\ket{\psi}$, 
we start with a guess of the MPS. The DMRG algorithm minimises the ground state energy in the space of the
MPS, where matrix elements are treated as variational parameters. The minimisation is performed by a sweeping 
procedure where the matrix state at a site is optimised at one time, keeping all others fixed, then optimising 
the next matrix, and so on. When all matrices are optimised, we again sweep back and forth until convergence 
is achieved. The convergence is guaranteed as the energy goes down at each iteration step.
%%%%%%%%%%%%%%%%%%%%%%%%%%%%%%%%%%%%%%%%%%%%%%%%
\begin{figure}[h!]
\centering
\includegraphics[width=0.35\textwidth]{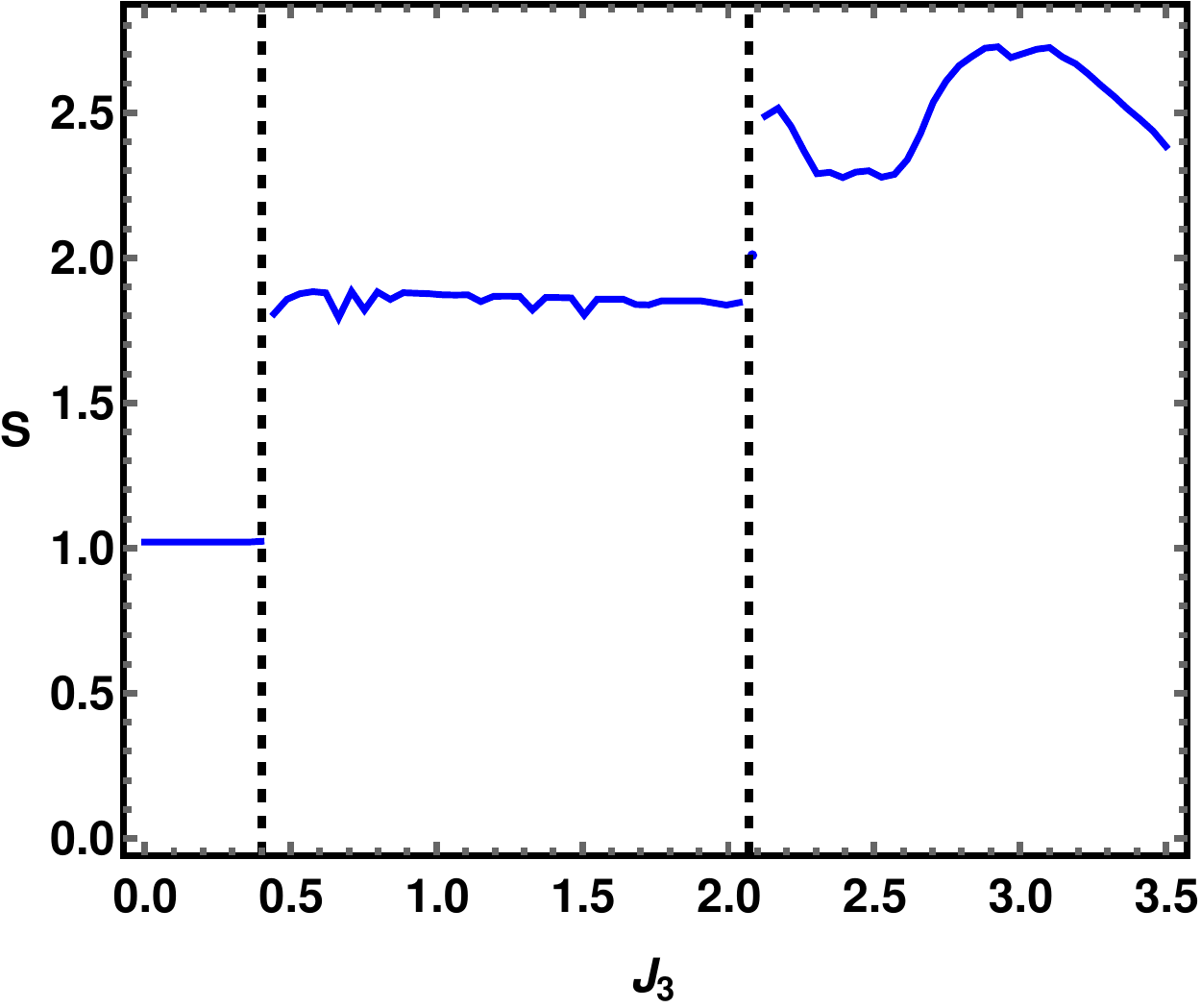}
\caption{In the absence of four-spin interactions, the Entanglement entropy $S$ as a f
unction of $J_{3}$ for $H=0$, $J_{1}=1.2$, $J_{2}=0.8$, $J_{13}=J_{23}=J_{3}$, and $J_{14}=J_{24}=0$.}
\label{entangle3spin}
\end{figure}
%%%%%%%%%%%%%%%%%%%%%%%%%%%%%%%%%%%%%%%%%%%%%%%%
 %%%%%%%%%%%%%%%%%%%%%%%%%%%%%%%%%%%%%%%%%%%%%%%%
\begin{figure}[h!]
\centering
\includegraphics[width=0.35\textwidth]{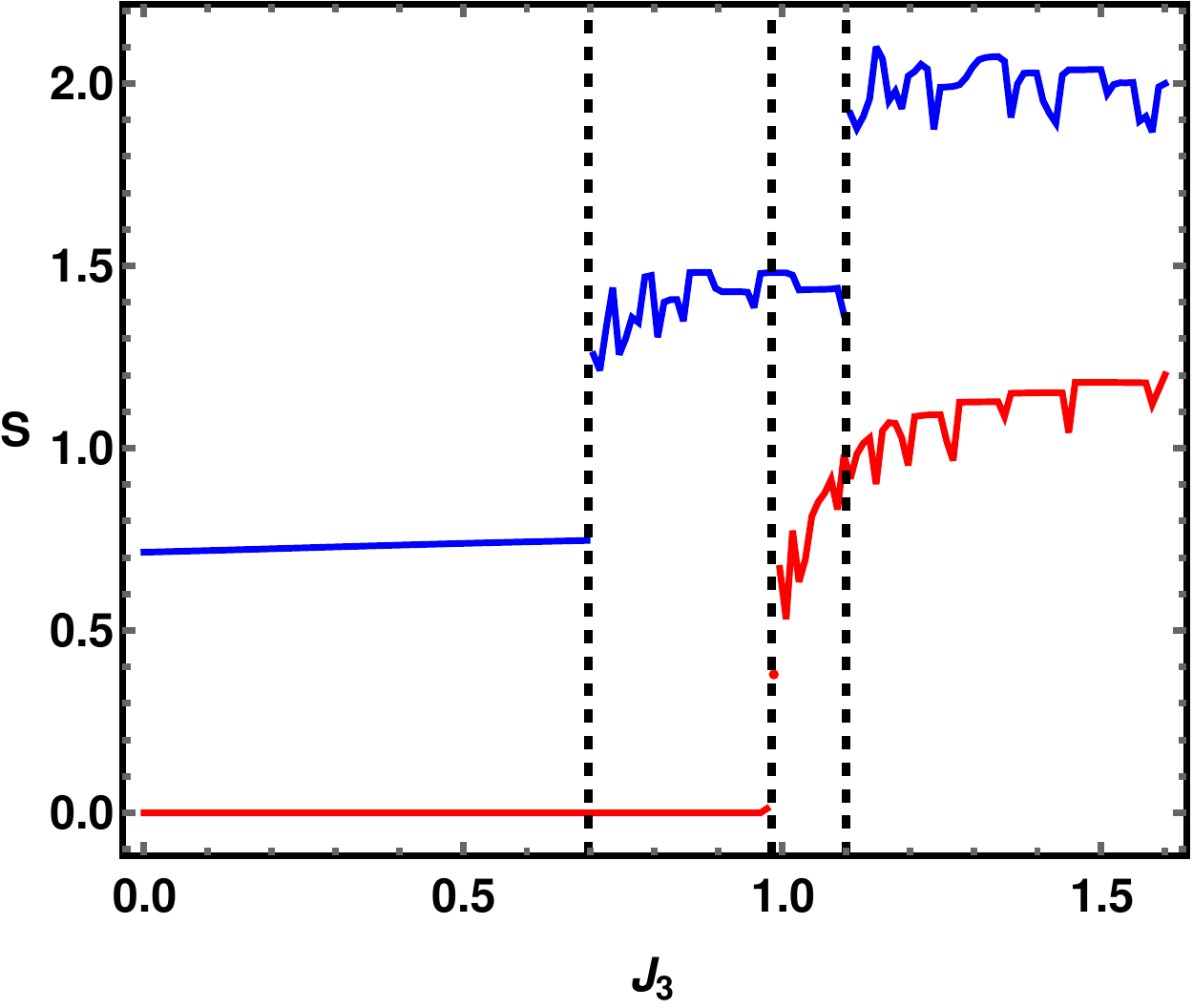}
\caption{The Entanglement entropy $S$ as a function of $J_{3}$ for the Hamiltonian of 
Eq. (\ref{parHamilt}). The blue and red curves correspond to $h=0.25$ and $h=2.5$ respectively, for fixed $J=1$.}
\label{entangle}
\end{figure}
%%%%%%%%%%%%%%%%%%%%%%%%%%%%%%%%%%%%%%%%%%%%%%%%
The DMRG uses the entanglement properties of a bipartite system to produce a state with minimal energy. 
In order to quantify the entanglement or quantum information, the most common measure is the Von Neumann entanglement 
entropy denoted by $S$. It is defined in a standard fashion, by bipartition of a composite system into two subsystems, $A$ and $B$,
\begin{equation}
S=-Tr\left[\rho_{A}\log\left(\rho_{A}\right)\right]=-Tr\left[\rho_{B}\log\left(\rho_{B}\right)\right]~,\label{EEeq}
\end{equation}
where $\rho_{A}$ and $\rho_{B}$ are the reduced density matrices of subsystems $A$ and $B$ respectively.

\subsection{EE with three-spin interactions}

We will numerically calculate the ground state EE in the simplest case $\mu_{1}=\mu_{2}=1$, $J_{13}=J_{23}=J_{3}$, 
and $J_{14}=J_{24}=0$ using DMRG method. We split the spin chain of $N$ cells at the centre into two subsystems. 
Using singular value decomposition, the MPS is divided into left and right bipartition. 
Then, the DMRG calculations are performed using the library \cite{itensor} on $N=51$ cells with maximum 
bond dimension $\chi=300$. The numerical results of EE as a function of $J_{3}$ is plotted in 
Fig. (\ref{entangle3spin}), for $H=0$, $J_{1}=1.2$, and $J_{2}=0.8$. When  $H=0$, there are two QPTs. 
One occurs at $J_{3}>(J_{1}-J_{2})=0.4$, which is the onset of the first-order QPT. Another occurs at 
$J_{3}=J_{1}+J_{2}=2$, when the line of second-order QPT, i.e., $H_{c_{2}}$ intersects the first-order 
QPT line ($J_{3}>0.4$, $H=0$). From Fig. (\ref{entangle3spin}), we see that the EE shows a 
sharp jump whenever the critical point is approached, irrespective of the order of the QPT.

\subsection{EE with four-spin interactions } 
Similar to the case of three-spin interactions, we have also plotted the numerical results of EE of our 
model Eq. (\ref{parHamilt}) as a function of  $J_{3}$ in Fig. (\ref{entangle}). The blue and red curves in 
Fig. (\ref{entangle}) correspond to $h=0.25$ and $h=2.5$, respectively, for a fixed value of $J=1$. 
When $h=0.25$, there are two critical points at $J_{3}=0.6958$ and $J_{3}=1.0958$. When $h=2.5$, the critical 
point is observed at $J_{3}=0.9753$. The EE clearly shows sharp jumps to indicate these second-order QPTs.
   
We conclude that for both the first and second-order QPTs, the EE shows a discontinuous nature when the parameters 
takes their critical values. For the first-order QPT, the result is in agreement with \cite{wu}, in which the 
first-order QPT is associated with the discontinuity in the density matrix elements. However, the discontinuity in 
EE for the second-order QPT in our case is contrary to what is observed in \cite{wu}, which suggests that the 
discontinuity or divergence should be present in the first derivative of EE for second-order QPTs, 
instead of EE itself. The reason for this result is that our model behaves differently from the other 
known exactly solvable spin chains. As we have mentioned before, there are two branches of the energy spectrum, 
${\cal E}_{k,1,2}$. Depending on the magnetic field $h$ and the three spin interaction parameter $J_{3}$, these 
branches completely or partially lie above or below the zero level, resulting in different magnetisations $m^{z}$. 
The different phases in the ground state are thus determined by the values of $m^{z}=0/1$ (antiferromagnetic/ferromagnetic) 
and $0<m^{z}<1$ (ferrimagnetic). Hence all the phase diagram characteristics of the model are determined by the signs 
of ${\cal E}_{k,1,2}$, which can be positive, negative, or zero. It is important to note that this behaviour is 
different from spin models with NN interactions, in which QPTs usually occur whenever the energy gap between the branches 
of energy spectra vanishes. For those models, the discontinuity is present in the derivative of EE for second-order QPTs, 
as concluded in \cite{wu}. But in our case, for different phases in the phase diagram, the ground state wavefunction 
$\ket{\Psi}$ takes distinct values, as the summation over the Fourier modes $k$ runs over specific ranges. The reduced density operator 
$\rho_{A,B}$ in Eq. (\ref{EEeq}) is the partial trace of the outer product of the ground state wavefunction 
$\ket{\Psi}$. So it becomes discontinuous whenever the phase change occurs, irrespective of the order of QPTs. 
The DMRG, a numerical technique for finding the ground state energy and wavefunction, confirms our analytical 
results of section $\Romannum{2}$ by showing the discontinuous nature of EE for both the first and second-order 
QPTs in Figs. (\ref{entangle3spin}) and (\ref{entangle}).

\section{Conclusions and discussions}
\label{conclusions}

In this paper, we have carried out a comprehensive analysis of information
theoretic geometry in a spin model with three and four spin interactions
along with an alternating NN coupling. In particular, we computed the
quantum information metric and the related Fubini-Study complexity, the
Nielsen complexity in static as well as quench scenarios and the entanglement
entropy in these models. The motivation and importance of this study lies
in the fact that the nature of phase transitions here are very different
from the usual ones in models with NN interactions, such as
the transverse field XY model, and it is important and interesting to
quantify information theoretic measures and contrast these with the ones
arising out of NN interactions. 

Here, we find a number of distinctive features, which are very different from ones 
in models with NN interactions. The derivative of the NC does
not diverge across a continuous phase transition, and the Fubini-Study complexity
also does not show any special behaviour across two ferrimagnetic phases. 
The entanglement entropy shows a discontinuity across both first and second
order phase transitions. These are in sharp contrast to the ones reported in
the literature, in models with only NN interactions. Our analysis points to the
fact that several features of information geometry that are valid in NN interaction
models cease to be valid when three and four spin interactions are included.

\appendix
\section{}
\label{AppA}
In this appendix, we list the transformations to diagonalise the Hamiltonian of Eq. (\ref{parHamilt}).

\noindent
(i) \textbf{Jordan-Wigner transformation}
\begin{eqnarray}
S_{n,1}^{z}&=&\dfrac{1}{2}\sigma_{n,1}^{z}=\dfrac{1}{2}-a_{n,1}^{\dagger}a_{n,1}, \notag\\
S_{n,2}^{z}&=&\dfrac{1}{2}\sigma_{n,2}^{z}=\dfrac{1}{2}-a_{n,2}^{\dagger}a_{n,2}, \notag\\
S_{n,1}^{+}&=&S_{n,1}^{x}+i S_{n,1}^{y}=\prod\limits_{m<n}\sigma_{m,1}^{z}\sigma_{m,2}^{z}a_{n,1}, \notag\\
S_{n,1}^{-}&=&S_{n,1}^{x}-i S_{n,1}^{y}=a_{n,1}^{\dagger}\prod\limits_{m<n}\sigma_{m,1}^{z}\sigma_{m,2}^{z}, \notag\\
S_{n,2}^{+}&=&\prod\limits_{m<n}\sigma_{m,1}^{z}\sigma_{m,2}^{z}\sigma_{n,1}^{z}a_{n,2},\notag\\
S_{n,2}^{-}&=&a_{n,2}^{\dagger}\prod\limits_{m<n}\sigma_{m,1}^{z}\sigma_{m,2}^{z}\sigma_{n,1}^{z}~,
\end{eqnarray}
where $a_{n,1,2}^{\dagger}$ and $a_{n,1,2}$ are creation and annihilation operators which satisfy 
usual anti-commutation relations. After the Jordan-Wigner transformation, we obtain
\begin{eqnarray}
{\cal H}&=&-\frac{h}{2}\sum\limits_{n}\left(2-3a_{n,1}^{\dagger}a_{n,1} -a_{n,2}^{\dagger}a_{n,2}\right)\notag\\& 
&-J\sum\limits_{n}\left(a_{n,2}^{\dagger}a_{n,1} +a_{n,1}^{\dagger}a_{n,2}\right)\notag\\& 
&+\frac{1}{2}\sum\limits_{n}\left(a_{n+1,1}^{\dagger}a_{n,2} +a_{n,2}^{\dagger}a_{n+1,1}\right)\notag\\& 
&-\frac{J_{3}}{4}\sum\limits_{n}\Big(5a_{n+1,1}^{\dagger}a_{n,1}+5a_{n,1}^{\dagger}a_{n+1,1}
+a_{n+1,2}^{\dagger}a_{n,2}+\notag\\& &a_{n,2}^{\dagger}a_{n+1,2}\Big)-\frac{1}{2}\sum\limits_{n}
\left(a_{n+1,2}^{\dagger}a_{n,1}+a_{n,1}^{\dagger}a_{n+1,2}\right)~.\notag\\
\end{eqnarray}
(ii) \textbf{Fourier transformation}
\begin{eqnarray}
a_{n,1}&=&\dfrac{1}{\sqrt{N}}\sum\limits_{k}d_{k,1}e^{i kn},\notag\\
a_{n,1}^{\dagger}&=&\dfrac{1}{\sqrt{N}}\sum\limits_{k}d_{k,1}^{\dagger}e^{-i kn}~,
\end{eqnarray}
and similarly for $a_{n,2}^{\dagger}$, and $a_{n,2}$. The Hamiltonian obtained after performing 
the Fourier transform is given by
\begin{eqnarray}
{\cal H}&=&\sum\limits_{k}\!\bigg[\!\!\left(\frac{3h}{2}-\frac{5J_{3}}{2}\cos k\right)\!d_{k,1}^{\dagger}d_{k,1}\notag\\
&+&\left(\frac{h}{2}-\frac{J_{3}}{2}\cos k\right)\!d_{k,2}^{\dagger}d_{k,2}\notag\\
& &-(J+i\sin k)d_{k,1}^{\dagger}d_{k,2}\notag\\
&-&(J-i\sin k)d_{k,2}^{\dagger}d_{k,1}-Nh\bigg]~,\label{hamiltfourier}
\end{eqnarray}
where with imposition of periodic boundary conditions, the quasi-momentum $k$ takes the values
\begin{center}
\(k=\frac{2\pi\lambda}{N}\),\, \(\lambda=-\frac{N-1}{2},....,-1,0,1,....,\frac{N-1}{2}\).
\end{center}
(iii) \textbf{Bogoliubov transformation}
\begin{eqnarray}
d_{k,1}&=&\frac{(J+i\sin k)}{\sqrt{J^{2}+\sin^{2}k}}\left(v_{k}b_{k,1}-u_{k}b_{k,2}\right),\notag\\
d_{k,2}&=&u_{k}b_{k,1}+v_{k}b_{k,2},\notag\\
d_{k,1}^{\dagger}&=&\frac{(J-i\sin k)}{\sqrt{J^{2}+\sin^{2}k}}\left(v_{k}b_{k,1}^{\dagger}-u_{k}b_{k,2}^{\dagger}\right),\notag\\
d_{k,2}^{\dagger}&=&u_{k}b_{k,1}^{\dagger}+v_{k}b_{k,2}^{\dagger}~.\label{bogoliubov}
\end{eqnarray}          
Here, the coefficients $u_{k}=\cos\left(\frac{\theta_{k}}{2}\right)$, $v_{k}=\sin\left(\frac{\theta_{k}}{2}\right)$, 
$\Lambda_{k}=\sqrt{\left(\frac{h}{2}-J_{3}\cos k\right)^{2}+J^{2}+\sin^{2}k}$, and 
$\cos \theta_{k}=(\frac{h}{2}-J_{3}\cos k)/\Lambda_{k}$.

\section{}
\label{AppB}
Here, we will provide the value of $k_{m}$ for which the critical line $h_{13}$ has been found 
out in Eq. (\ref{critical mag2}). This reads
\begin{equation}
k_m = -2 \tan ^{-1}\left(\sqrt{\frac{a+b}{c}}\right)~,
\end{equation}
where we define
\begin{eqnarray}
a&=&-64 J^2 J_{3}^2+5 J_{3}^4-120
   J_{3}^2-48\nonumber\\
b&=&16\sqrt{\left(J^2+1\right) \left(56
   J_{3}^4+48 J_{3}^2-5 J_{3}^6\right)}\nonumber\\
   c&=&64 J^2 J_{3}^2+5 J_{3}^4+8
   J_{3}^2-48~.
\end{eqnarray} 
\twocolumngrid


\begin{thebibliography}{999}
\bibitem{zanardi}
P.~Zanardi, P.~Giorda and M.~Cozzini, Phys. Rev. Lett. {\bf 99},
100603 (2007).
\bibitem{gu}
S.~Gu, Int. J. Mod. Phys. B {\bf 24}, 4371 (2010).
\bibitem{polkov}
M.~Kolodrubetz, V.~Gritsev and A.~Polkovnikov, Phys. Rev. B
{\bf 88}, 064304 (2013).
\bibitem{Chapman}
S.~Chapman, M.~P.~Heller, H.~Marrochio and F.~Pastawski,
Phys. Rev. Lett. {\bf 120}, 121602 (2018).
\bibitem{Nielsen}
M.~A.~Nielsen, M.~R.~Dowling, M.~Gu and A.~C.~Doherty, Science {\bf 311}, 1133 (2006).
\bibitem{Nielsen1}
M.~A.~Nielsen, arXiv:quant-ph/0502070;
M.~R.~Dowling and M.~A.~Nielsen, arXiv:quant-ph/0701004.
\bibitem{Peres}
A.~Peres, Phys. Rev. A {\bf 30}, 1610 (1984).
\bibitem{Hamma1}
J.~Happola, G.~B.~Halasz, and A.~Hamma, Phys. Rev. A {\bf 85}, 032114 (2012).
\bibitem{osborne}
T.~J.~Osborne and M.~A.~Nielsen, Phys.\ Rev.\ A {\bf 66},
032110 (2002).
\bibitem{oster}
A.~Osterloh, L.~Amico, G.~Falci and R.~Fazio, Nature {\bf 416}, 608 (2002).
\bibitem{vidal}
G.~Vidal, J.~I.~Latorre, E.~Rico and A.~Kitaev, Phys.\ Rev.\ Lett.\ {\bf 90}, 227902 (2003).
\bibitem{su}
S.~Q.~Su, J.~L.~Song and S.~J.~Gu, Phys.\ Rev.\ A {\bf 74},
032308 (2006).
\bibitem{Susskind1}
L. Susskind, Fortsch. Phys. {\bf 64} 44 (2016).
\bibitem{Susskind2}
L. Susskind, Fortsch. Phys. {\bf 64} 49 (2016).
\bibitem{Susskind3}
A.R. Brown, D.A. Roberts, L. Susskind, B. Swingle and Y. Zhao, 
Phys. Rev. Lett. {\bf 116} 191301 (2016).
\bibitem{Susskind4}
A.R. Brown, D.A. Roberts, L. Susskind, B. Swingle and Y. Zhao, Phys. Rev. {D 93} 086006 (2016). 
\bibitem{Myers1}
R.~Jefferson and R.~C.~Myers,
%``Circuit complexity in quantum field theory,''
JHEP {\bf 1710}, 107 (2017).
\bibitem{BSS}
A. Bhattacharyya, A. Shekar, and A. Sinha, JHEP \textbf{10}, 140 (2018).
\bibitem{Myers2}
M.~Guo, J.~Hernandez, R.~C.~Myers and S.~M.~Ruan,
%``Circuit Complexity for Coherent States,''
JHEP {\bf 1810}, 011 (2018).
\bibitem{KKS}
R. Khan, C. Krishnan, and S. Sharma, Phys. Rev. D \textbf{98}, 126001 (2018).
{\bibitem{HM}
L. Hackl, and R. C. Myers  JHEP \textbf{07} (2018), 139.}
\bibitem{ABHKM}
T. Ali, A. Bhattacharyya, S.S. Haque, E.H. Kim, and N. Moynihan, JHEP {\bf 04} (2019) 087.
\bibitem{tapo}
N.~Jaiswal, M.~Gautam, and T.~Sarkar, Phys.\ Rev.\ E {\bf 104}, 024127 (2021).
\bibitem{tapo1}
N.~Jaiswal, M.~Gautam, and T.~Sarkar,
arXiv:2110.02099 [quant-ph].
\bibitem{GGCHV}
D. Gutierrez-Ruiz, D. Gonzalez, J. Chavez-Carlos, J. Hirsch, J. D. Vergara, Phys. Rev. B \textbf{103} 174104 (2021).
\bibitem{tapo2}
K.~Pal, K.~Pal and T.~Sarkar,
%``Complexity in the Lipkin-Meshkov-Glick Model,''
[arXiv:2204.06354 [quant-ph]].
\bibitem{Zvyagin}
A.~A.~Zvyagin and G.~A.~Skorobagat'ko, Phys.\ Rev.\ B {\bf 73}, 024427 (2006).
\bibitem{Zvyagin1}
A.~A.~Zvyagin and V. O. Cheranovskii, Low Temp. Phys. {\bf 35}, 6 (2009).
\bibitem{Zvyaginbook}
A.~A.~Zvyagin, Finite size effects in correlated electron models: exact results, World Scientific, (Imperial College Press,  2005).
\bibitem{white}
S.~R.~White, Phys.\ Rev.\ Lett.\ {\bf 69}, 2863 (1992); S.~R.~White, Phys.\ Rev.\ B {\bf 48}, 10345 (1993).
\bibitem{Scholl}
U.~Schollwoeck, Rev. Mod. Phys. {\bf 77} (2005);
U.~Schollwoeck, Ann. Phys. {\bf 326}, 96-192, (2011).
\bibitem{osborne1}
T.~J.~Osborne and M.~A.~Nielsen, Quantum Inf. Process, {\bf 1},
45-53 (2002).
\bibitem{deger}
A.~Deger, T.~C.~Wei, Quantum Inf. Process, {\bf 18}, 326 (2019).
\bibitem{huang}
H.~H.-Lin, Commun. Theor. Phys. {\bf 55}, 349 (2011).
\bibitem{Liu}
F.~Liu, S.~Whitsitt, J.~B.~Curtis, R.~Lundgren, P.~Titum, Z.~C.~Yang, J.~R.~Garrison and A.~V.~Gorshkov,
%``Circuit complexity across a topological phase transition,''
Phys.\ Rev.\ Res.\  {\bf 2}, 013323 (2020).
\bibitem{ZanLE}
H. T. Quan, Z. Song, X. F. Liu, P. Zanardi, and C. P. Sun,
Phys. Rev. Lett. {\bf 96}, 140604 (2006).
\bibitem{itensor}
M.~Fishman, S.~R.~White, and E.~M.~Stoudenmire, arXiv:2007.14822,  (2020).
\bibitem{wu}
L.~A.~Wu, M.~S.~Sarandy, and D.~A.~Lidar, Phys.\ Rev.\ Lett. {\bf 93}, 250404 (2004).
\bibitem{DG}
G.~Di Giulio and E.~Tonni,
%``Subsystem complexity after a global quantum quench,''
JHEP \textbf{05}, 022 (2021).
\bibitem{tapo5}
K.~Pal, K.~Pal, A.~Gill and T.~Sarkar,
%``Evolution of circuit complexity in a harmonic chain under multiple quenches,''
[arXiv:2206.03366 [quant-ph]].
\end{thebibliography}
\end{document}